\documentclass[aps,prb,twocolumn,superscriptaddress,groupedaddress,10pt]{revtex4}
\usepackage{amsmath}
\usepackage{amssymb}
\usepackage{graphicx}
\usepackage{epsfig}
\usepackage{dcolumn}
\usepackage{bm}
\usepackage{bm}
\usepackage{blindtext}
\usepackage{natbib}
\usepackage{siunitx} 

\usepackage{float}

\usepackage[titletoc]{appendix}

\usepackage{simplewick}

\usepackage{bbm}
\makeatletter
\newcommand{\mathleft}{\@fleqntrue\@mathmargin0pt}
\newcommand{\mathcenter}{\@fleqnfalse}
\makeatother

\usepackage{afterpage}

\usepackage[usenames,dvipsnames]{xcolor}
\usepackage{amsthm}
\usepackage{booktabs}
\usepackage{hyperref}
\usepackage{tikz}
\usetikzlibrary{calc}
\usepackage[printwatermark]{xwatermark}
\newcommand{\tikzmark}[1]{\tikz[overlay,remember picture] \node (#1) {};}
\newcommand{\DrawBox}[3][]{%
    \tikz[overlay,remember picture]{
    \draw[black,#1]
      ($(#2)+(-0.1em,2.0ex)$) rectangle
      ($(#3)+(0.1em,-0.5ex)$);}
}

\usepackage{mathtools}

\def\be{\begin{equation}} \def\ee{\end{equation}}
\def\bea{\begin{eqnarray}} \def\eea{\end{eqnarray}}

\def\nn{\nonumber}

\newcommand{\ket}[1]{| #1 \rangle}
\newcommand{\bra}[1]{\langle #1 |}

\begin{document}
\title{
Classical spin order near antiferromagnetic  Kitaev point  in the spin-1/2 Kitaev-Gamma chain
}

\author{Wang Yang}
\affiliation{Department of Physics and Astronomy and Stewart Blusson Quantum Matter Institute,
University of British Columbia, Vancouver, B.C., Canada, V6T 1Z1}

\author{Alberto Nocera}
\affiliation{Department of Physics and Astronomy and Stewart Blusson Quantum Matter Institute, 
University of British Columbia, Vancouver, B.C., Canada, V6T 1Z1}

\author{Erik S. S{\o}rensen}
\affiliation{Department of Physics and Astronomy, McMaster University 1280 Main St. W., Hamilton ON L8S 4M1, Canada.}

\author{Hae-Young Kee}
\affiliation{Department of Physics, University of Toronto, Ontario M5S 1A7, Canada}
\affiliation{Canadian Institute for Advanced Research/Quantum Materials Program, Toronto, Ontario MSG 1Z8, Canada}

\author{Ian Affleck}
\affiliation{Department of Physics and Astronomy and Stewart Blusson Quantum Matter Institute, 
University of British Columbia, Vancouver, B.C., Canada, V6T 1Z1}

\begin{abstract}

A minimal Kitaev-Gamma model has been recently investigated to understand various Kitaev systems.
In  the one-dimensional Kitaev-Gamma chain, an emergent SU(2)$_1$ phase and a rank-1 spin ordered phase with $O_h\rightarrow D_4$ symmetry breaking were identified using non-Abelian bosonization and numerical techniques.  
However, puzzles near the antiferromagnetic Kitaev region with finite Gamma interaction remained unresolved. 
Here we focus on this parameter region and find that there are two new phases,
namely, a rank-1 ordered phase with an $O_h\rightarrow D_3$ symmetry breaking, and a peculiar Kitaev phase. 
Remarkably, the $O_h\rightarrow D_3$ symmetry breaking corresponds to the classical magnetic order, but appears in a region very close to the antiferromagnetic Kitaev point where the quantum fluctuations are presumably very strong.
In addition, a two-step symmetry breaking $O_h\rightarrow D_{3d}\rightarrow D_3$ is numerically observed as the length scale is increased: 
At short and intermediate length scales, the system behaves as having a rank-2 spin nematic order with $O_h\rightarrow D_{3d}$ symmetry breaking;
and at long distances,  time reversal symmetry is further broken  leading to the $O_h\rightarrow D_3$ symmetry breaking. 
Finally, there is no numerical signature of spin orderings nor Luttinger liquid behaviors in the Kitaev phase whose nature is worth further studies.

\end{abstract}
\maketitle

\section{Introduction}

Low dimensional quantum magnetism is among the most active research areas in modern condensed matter physics \cite{Fazekas1999,Lauchli2006}.
The interplays between quantum fluctuations, low dimensionality and frustrations lead to exotic magnetic properties including various magnetic orderings, topological orders and spin liquid behaviors \cite{Balents2010,Witczak-Krempa2014,Rau2016,Savary2017,Winter2017,Zhou2017}.
The strong spin-orbit coupling effects in $4d$- and $5d$-electron compounds have added to the richness of the strongly correlated magnetic behaviors. 
Examples of this kind include the Kitaev materials on the two-dimensional (2D) honeycomb lattice, 
which are proposed to host exotic fractionalized excitations including Majorana fermions and nonabelian anyons \cite{Kitaev2006,Nayak2008,Jackeli2009,Chaloupka2010,Singh2010,Price2012,Singh2012,Plumb2014,Kim2015,Baek2017,Leahy2017,Sears2017,Wolter2017,Zheng2017,Kasahara2018}. 
A generalized Kitaev model containing symmetry allowed terms  in addition to the Kitaev interaction
have been proposed and analyzed to described the real Kitaev materials \cite{Rau2014,Catuneanu2018,Gohlke2018,Ran2017,Wang2017}.

Recently, the generalized Kitaev spin-1/2 model has also been actively studied in one-dimensional (1D) systems
 \cite{Agrapidis2018,Yang2019,Yang2020a,Luo2020,You2020},
which may be realized in Ruthenium stripes in the RuCl$_3$ materials \cite{Gruenewald2017}.
The spin ladder cases have also been investigated \cite{Agrapidis2019,Catuneanu2019,Sorensen2021}. 
In Ref. \onlinecite{Yang2019}, the phase diagram of the 1D Kitaev-Gamma spin-1/2 chain has been studied.
It is shown that about $67\%$ of the phase diagram of the Kitaev-Gamma chain is described by an emergent SU(2)$_1$ Wess-Zumino-Witten (WZW) model, 
and besides this, an ordered phase of rank-1 spin orders with an $O_h\rightarrow D_4$ symmetry breaking is identified, where $O_h$ is the full octahedral group and $D_n$ represents the dihedral group of order $2n$.

\begin{figure}
\center
\includegraphics[width=7cm]{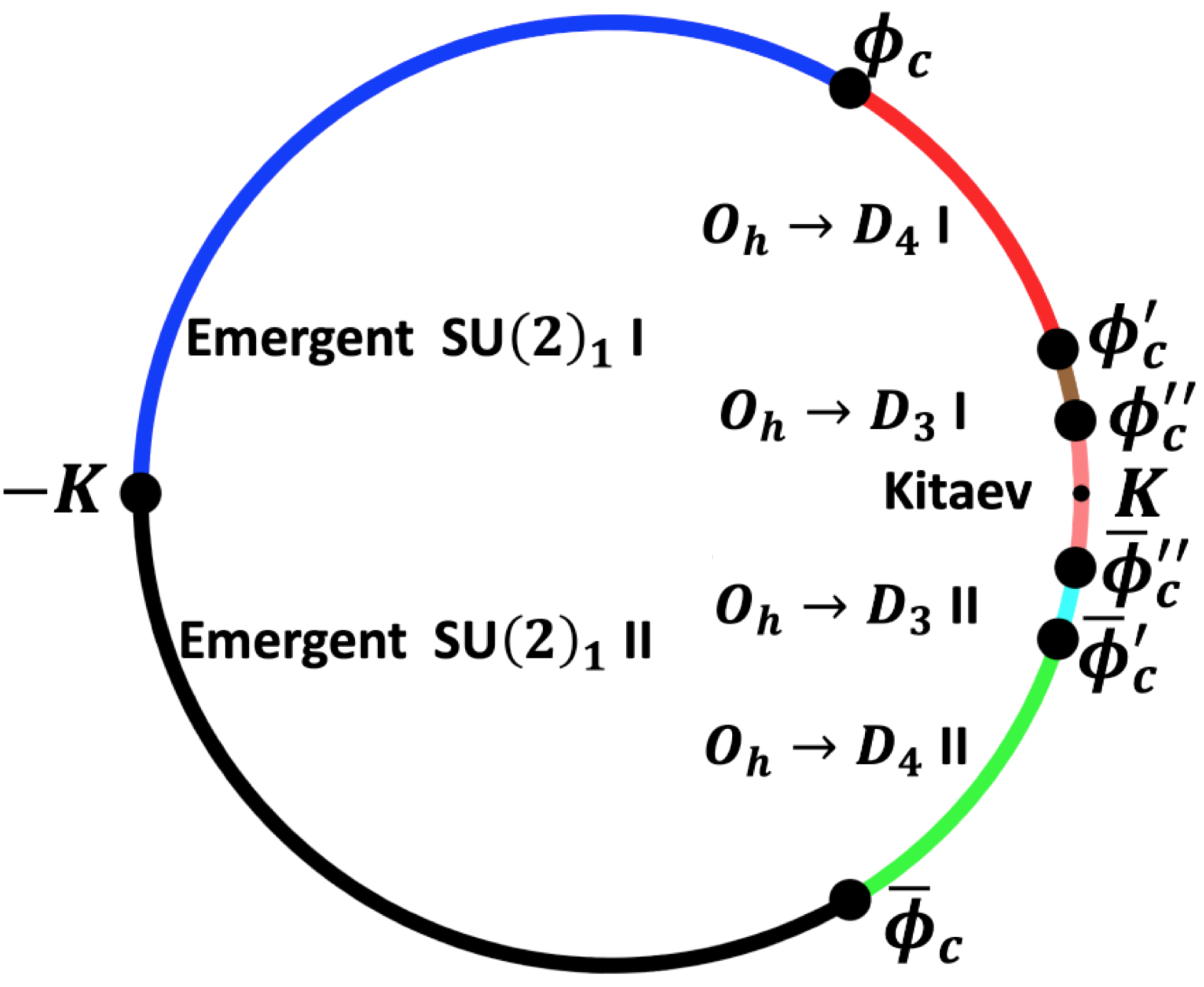}
\caption{Phase diagram of the spin-1/2 Kitaev-Gamma chain,
in which $K=\cos(\phi)$ and $\Gamma=\sin(\phi)$.
The phases marked with ``I" and ``II" are related by  the equivalence $(K,\Gamma)\cong (K,-\Gamma)$.
} 
\label{fig:phase_diagram}
\end{figure}

In this work, we focus on the unresolved phases near the antiferromagnetic (AFM) Kitaev region.
To elaborate our current study, we begin with a quick review of the phase diagram  of the Kitaev-Gamma chain shown in Fig. \ref{fig:phase_diagram}.
Since changing the sign of the Gamma coupling leads to an equivalent Hamiltonian \cite{Yang2019},
it is enough to consider the upper half circle in Fig. \ref{fig:phase_diagram}  where all the phases are numbered by ``I" except the ``Kitaev" phase.
By a combination of symmetry analysis, density matrix renormalization group (DMRG), infinite DMRG  (iDMRG),  and exact diagonalization (ED) numerical methods, 
two new phases are identified, namely the ``$O_h\rightarrow D_3$ I" and ``Kitaev" phases,
in addition to the already established ``emergent SU(2)$_1$ I" and ``$O_h\rightarrow D_4$ I" phases \cite{Yang2019}.

The $O_h\rightarrow D_3$ symmetry breaking corresponds to the classical spin order as discussed in Ref. \onlinecite{Yang2020} where the spin-$S$ Kitaev-Gamma ($K$-$\Gamma$) chain  is considered for $S\geq 1$,
occupying  the region around the point $K=\Gamma$.
Counterintuitively, in the  spin-1/2 case, this classical order appears in a very narrow region  around the infinitely degenerate AFM  Kitaev point \cite{Brzezicki2007} where quantum fluctuations are presumably strong.
In addition, a two-step symmetry breaking $O_h\rightarrow D_{3d}\rightarrow D_3$ is numerically observed as the length scale is increased: 
At short and intermediate length scales, the system behaves as having a rank-2 spin nematic order with $O_h\rightarrow D_{3d}$ symmetry breaking,
where $D_{3d}\cong D_3\times \mathbb{Z}_2^T$
in which $\mathbb{Z}_2^T$ is the $\mathbb{Z}_2$ group generated by the time reversal operation;
and at long distances,  time reversal symmetry is further broken  leading to the $O_h\rightarrow D_3$ symmetry breaking. 
Finally, the nature of the ``Kitaev" phase in Fig. \ref{fig:phase_diagram}  remains unclear with no numerical evidence of magnetic orderings nor Luttinger liquid behaviors.
Whether more exotic  orderings like the topological string order \cite{Catuneanu2019} exist in the ``Kitaev" phase is worth further studies.

The rest of the paper is organized as follows. 
In Section \ref{sec:hamiltonian}, the model Hamiltonian is presented and the symmetries of the model  are analyzed. 
Section \ref{sec:orders} summarizes the magnetic orders discussed in this paper.
In Section \ref{sec:spin_nematic}, the spin-nematic order at short and intermediate length scales is discussed.
Section \ref{sec:emerge_rank1} is devoted to a thorough discussion of the classical $O_h\rightarrow D_3$ order emerging  at long distances with  plenty of numerical evidence. 
Section \ref{sec:origin_classical} proposes an argument for the possible origin of the classical order in the region with presumably strong quantum fluctuations.
In Section \ref{sec:Kitaev}, the peculiar Kitaev phase is discussed.
Finally in Sec. \ref{sec:concl},  the main results and open questions of the paper are briefly summarized.

\section{Model Hamiltonian and symmetries}
\label{sec:hamiltonian}

In this section, we first present the model Hamiltonian, and then briefly review the six-sublattice rotation and the symmetries of the system.

\subsection{Model Hamiltonian}

\begin{figure}
\includegraphics[width=8.5cm]{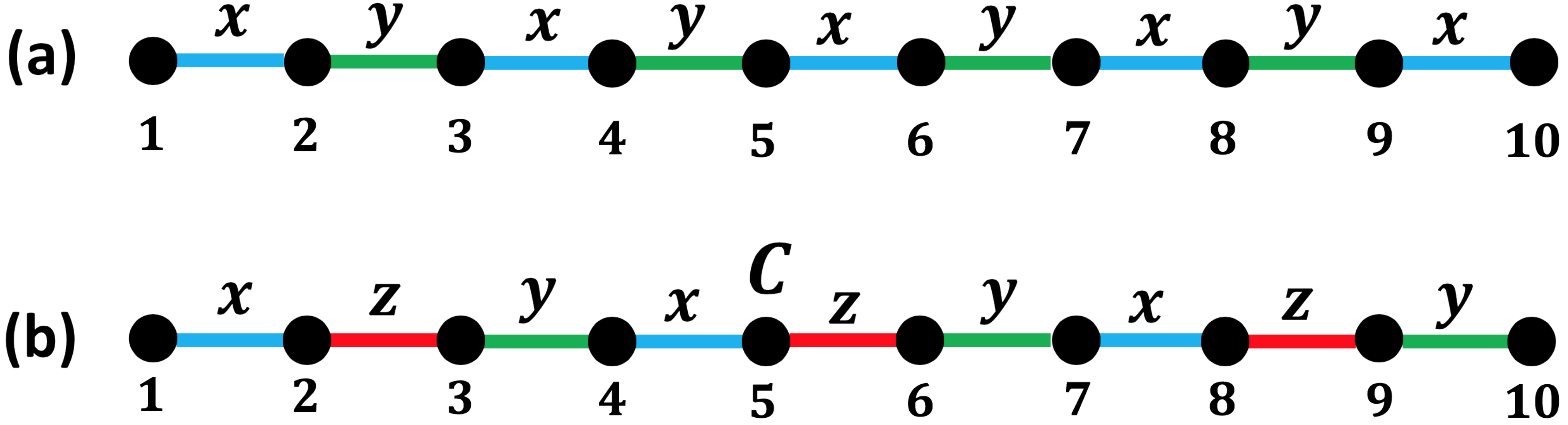}
\caption{Bond structures in (a) the unrotated and (b) the six-sublattice rotated frames.
}
\label{fig:bonds}
\end{figure}

The Hamiltonian of the spin-1/2 Kitaev-Gamma chain is defined as \cite{Yang2019}
\bea
H=\sum_{<ij>\in\gamma\,\text{bond}}\big[ KS_i^\gamma S_j^\gamma+\Gamma (S_i^\alpha S_j^\beta+S_i^\beta S_j^\alpha)\big],
\label{eq:H_unrot}
\eea
in which $i,j$ are two sites of nearest neighbors;
$\gamma=x,y$ is the spin direction associated with the $\gamma$ bond shown in Fig. \ref{fig:bonds} (a);
$\alpha\neq\beta$ are the two remaining spin directions other than $\gamma$;
$K$ and $\Gamma$ are the Kitaev and Gamma couplings, respectively.
In what follows, the two couplings will be parametrized as
\bea
K=\cos(\phi),~\Gamma=\sin(\phi),
\eea
where $\phi\in[0,2\pi]$.
Under a global spin rotation around the $z$-axis by $\pi$, i.e.,  $R(\hat{z},\pi):(S_i^x,S_i^y,S_i^z)\rightarrow (S_i^y,-S_i^x,S_i^z)$, 
the Kitaev term remains the same whereas $\Gamma$ changes to $-\Gamma$. 
Therefore, there is the equivalence
\bea
(K,-\Gamma) \cong (K,\Gamma),
\label{eq:equiv}
\eea 
or $\phi\cong 2\pi-\phi$.
Due to this equivalence, the phase diagram can be restricted to the parameter range  $\phi\in(0,\pi)$,
and in subsequent discussions, we will drop the numbering ``I" in the names of the phases for simplicity.

\subsection{The six-sublattice rotation}

A useful transformation $U_6$ with a periodicity of six sites is defined as \cite{Stavropoulos2018,Yang2019}
\bea
\text{Sublattice $1$}: & (x,y,z) & \rightarrow (x^{\prime},y^{\prime},z^{\prime}),\nn\\ 
\text{Sublattice $2$}: & (x,y,z) & \rightarrow (-x^{\prime},-z^{\prime},-y^{\prime}),\nn\\
\text{Sublattice $3$}: & (x,y,z) & \rightarrow (y^{\prime},z^{\prime},x^{\prime}),\nn\\
\text{Sublattice $4$}: & (x,y,z) & \rightarrow (-y^{\prime},-x^{\prime},-z^{\prime}),\nn\\
\text{Sublattice $5$}: & (x,y,z) & \rightarrow (z^{\prime},x^{\prime},y^{\prime}),\nn\\
\text{Sublattice $6$}: & (x,y,z) & \rightarrow (-z^{\prime},-y^{\prime},-x^{\prime}),
\label{eq:6rotation}
\eea
in which "Sublattice $i$" ($1\leq i \leq 6$) denotes the sites $i+6n$ ($n\in \mathbb{Z}$), and $S^\alpha$ ($S^{\prime \alpha}$) is abbreviated as $\alpha$ ($\alpha^\prime$) for short ($\alpha=x,y,z$).
After the six-sublattice rotation, the Hamiltonian $H^\prime=U_6 H U_6^{-1}$ acquires the form
\bea
H^\prime&=\sum_{<ij>\in \gamma\,\text{bond}}\big[ -KS_i^\gamma S_j^\gamma-\Gamma (S_i^\alpha S_j^\alpha+S_i^\beta S_j^\beta)\big],
\label{eq:6rotated}
\eea
in which the bond $\gamma=x,z,y$ is periodic in three sites as shown in Fig. \ref{fig:bonds} (b),
and the  prime has been dropped in $\vec{S}_i^\prime$ for simplicity.
The explicit expression of $H^\prime$ in Eq. (\ref{eq:6rotated}) is given in Appendix \ref{sec:Ham}.

We will stick to the six-sublattice rotated frame from here on in the remaining parts of this work unless otherwise stated.
The Hamiltonian is simplified in the six-sublattice rotated frame in the sense that there is no cross term $S_i^\alpha S_{i+1}^\beta$ where $\alpha\neq \beta$.
In particular, $H^\prime$ becomes SU(2) symmetric when $K=\Gamma$.
Due to the equivalence established in Eq. (\ref{eq:equiv}), the system also has hidden SU(2) symmetry at $K=-\Gamma$.
In the range $\phi\in[0,\pi]$, the points $\phi=\pi/4$ and $3\pi/4$ corresponds to an ferromagnetic (FM) and AFM Heisenberg model, respectively.

\subsection{The symmetry group}

In this section, the symmetry group of $H^\prime$ will be briefly reviewed which has been discussed in detail in Ref. \onlinecite{Yang2019,Yang2020}.

The Hamiltonian $H^\prime$ in Eq. (\ref{eq:6rotated}) is invariant under the 
time reversal operation $\mathcal{T}$, the screw operation $R_aT_a$, the coupled operation 
$R_II$, and the global spin rotations $R(\hat{\alpha},\pi)$ ($\alpha=x,y,z$),
in which: 
$T_a$ and $I$ represent the spatial translation by one site and the inversion around the point $C$ in Fig. \ref{fig:bonds} (b), respectively;
$R_a$ and $R_I$ are given by $R_a=R(\hat{n}_a,-2\pi/3)$ and $R_I=R(\hat{n}_I,\pi)$,
where $R(\hat{n},\theta)$ represents a global spin rotation around the $\hat{n}$-axis by an angle $\theta$, and the rotation axes $\hat{n}_a$, $\hat{n}_I$  are given by $\hat{n}_a=\frac{1}{\sqrt{3}}(1,1,1)^T$, $\hat{n}_I=\frac{1}{\sqrt{2}}(1,0,-1)^T$.
These symmetry operations generate the symmetry group $G$ of the system:
\bea
G=\mathopen{<}  \mathcal{T},R_aT_a,R_I I, R(\hat{x},\pi),R(\hat{y},\pi),R(\hat{z},\pi) \mathclose{>}.
\eea
Notice that the spatial translation by three sites $T_{3a}=(R_aT_a)^3$ is a group element of $G$, which generates an abelian normal subgroup. 
It has been shown in Ref. \onlinecite{Yang2019} that the quotient group $G/\mathopen{<}T_{3a}\mathclose{>}$ is isomorphic to $O_h$, 
where $O_h$ is the full octahedral group. 
Therefore, the group structure of $G$ can be represented as
\bea
G\cong O_h \ltimes 3\mathbb{Z},
\eea
where $3\mathbb{Z}$ is $\mathopen{<}T_{3a}\mathclose{>}$ for short, and $\ltimes$ is the semi-direct product.

\section{Phase diagram and magnetic orders}
\label{sec:orders}

\subsection{Phase diagram}

Here we give a brief summary about the phase diagram.
In Ref. \onlinecite{Yang2019}, the region $\phi\in[\phi_c,\pi]$ is shown to be a gapless phase described by an emergent SU(2)$_1$ WZW model,
where $\phi_c\simeq 0.33\pi$.
Also, a conventional rank-1 spin ordered phase with an $O_h\rightarrow D_4$ symmetry breaking has been identified within  $\phi\in[\phi_c^\prime,\phi_c]$ where $\phi_c^\prime \simeq 0.10\pi$.
However, the phase below $\phi_c^\prime$ was not studied in Ref. \onlinecite{Yang2019}.

In this work, we identify the region $\phi\in[\phi_c^{\prime\prime},\phi_c^\prime]$ to have a
classical spin order, where $\phi_c^{\prime\prime}\simeq 0.034\pi$ as determined from iDMRG calculations  discussed in Sec. \ref{sec:Kitaev}.
The symmetry breaking pattern is $O_h\rightarrow D_{3}$, 
 and the ground state degeneracy is $|O_h/D_3|=8$.
 The classical $O_h\rightarrow D_3$ order  has been discussed in detail in Ref. \onlinecite{Yang2020} for the spin-$S$ Kitaev-Gamma chain in the large-$S$ limit.
 However, in Ref. \onlinecite{Yang2020} where $S\geq 1$, the classical order is found to locate around $\phi=\pi/4$ which is occupied by the $O_h\rightarrow D_4$ phase in the spin-$1/2$ case\cite{Yang2019}.
 For the spin-1/2 Kitaev-Gamma chain, the classical order appears in the  vicinity of the infinitely degenerate \cite{Brzezicki2007} Kitaev point $K=0$ where quantum fluctuations are presumably strong. 
In addition, we will demonstrate that the system behaves like a spin-nematic order at short and intermediate length scales, and transits into the classical order only at sufficiently long distances.
Finally, the region $\phi\in[0,\phi_c^{\prime\prime}]$ is a different phase denoted as the ``Kitaev" phase in Fig. \ref{fig:phase_diagram}  whose nature  remains unclear.

Here we make some comments about the DMRG and iDMRG numerics that we have performed in the calculations. 
In our work, the DMRG method\cite{White1992,White1993} was used on chains with length up to $L=96$ sites and periodic boundary conditions. 
Even though it is known that DMRG convergence is hard for periodic boundary conditions,  we checked that our results are converged using up to $m = 1200$ states with a truncation error below $10^{-6}$ as in previous investigations in Ref. \onlinecite{Yang2019}.
For iDMRG, the unit cell size is chosen as $24$, and the bond dimension is $1000$.
The typical truncation error is $10^{-10}$.

\subsection{The classical $O_h\rightarrow D_3$ order}
\label{sec:OhD3}

\begin{figure}
\center
\includegraphics[width=7cm]{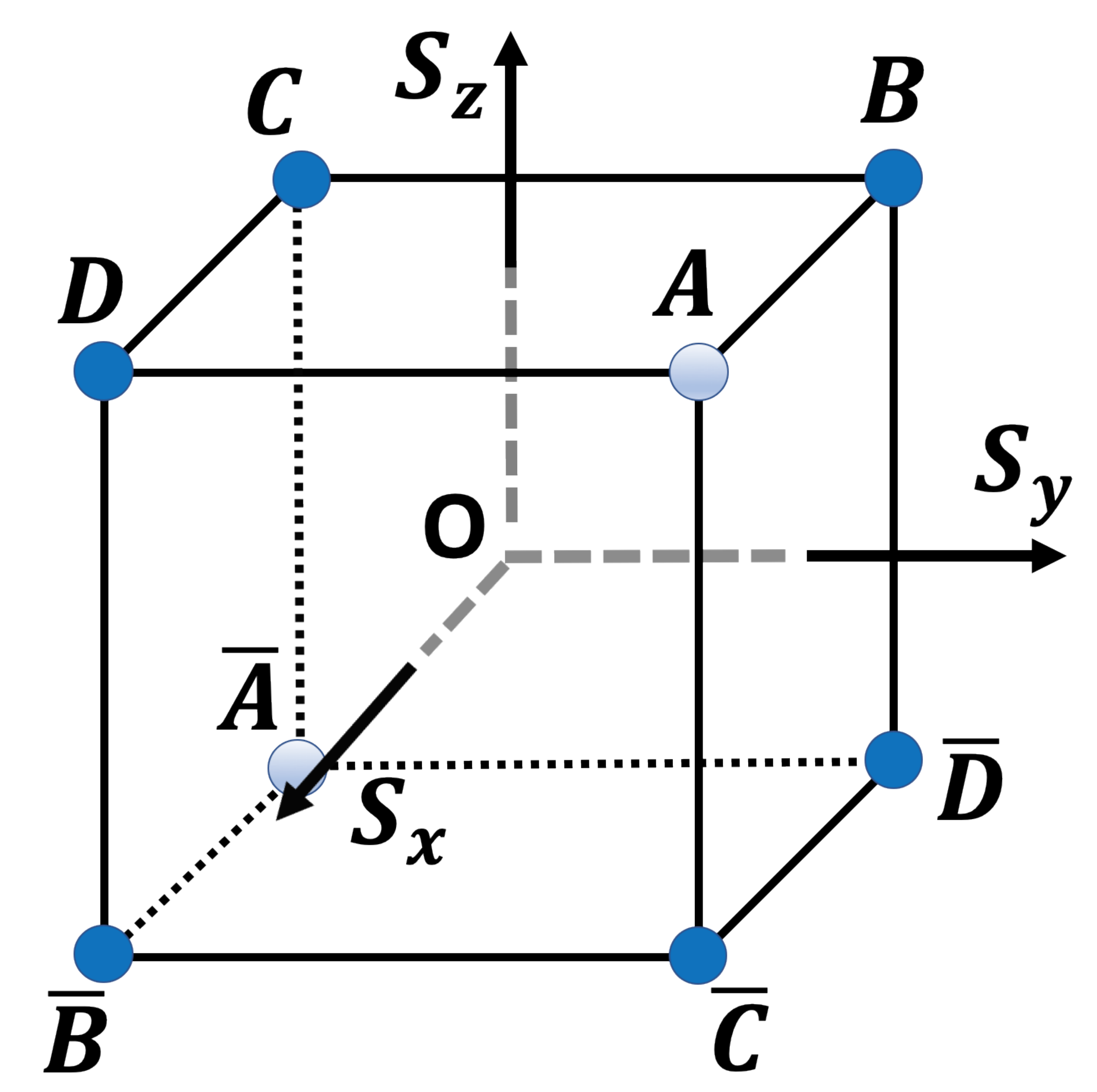}
\caption{``Center of mass" directions of the three spins within a unit cell 
in the six-sublattice rotated frame as represented by
the eight solid blue circles  for the eight degenerate ground states in the ``$O_h\rightarrow D_3$" phase.
The ground states of the two solid light blue circles (along the $\pm(1,1,1)$-directions) correspond to N\'eel ordering in the original frame. 
On the other hand, the states of the six solid dark blue circles exhibit six-site periodicities in the original frame.
The convention  of the coordinates is taken such that  the eight vertices of the cube are  located at $(\pm 1,\pm1,\pm1)$.
} 
\label{fig:spin_orders}
\end{figure}

\begin{figure}
\center
\includegraphics[width=8cm]{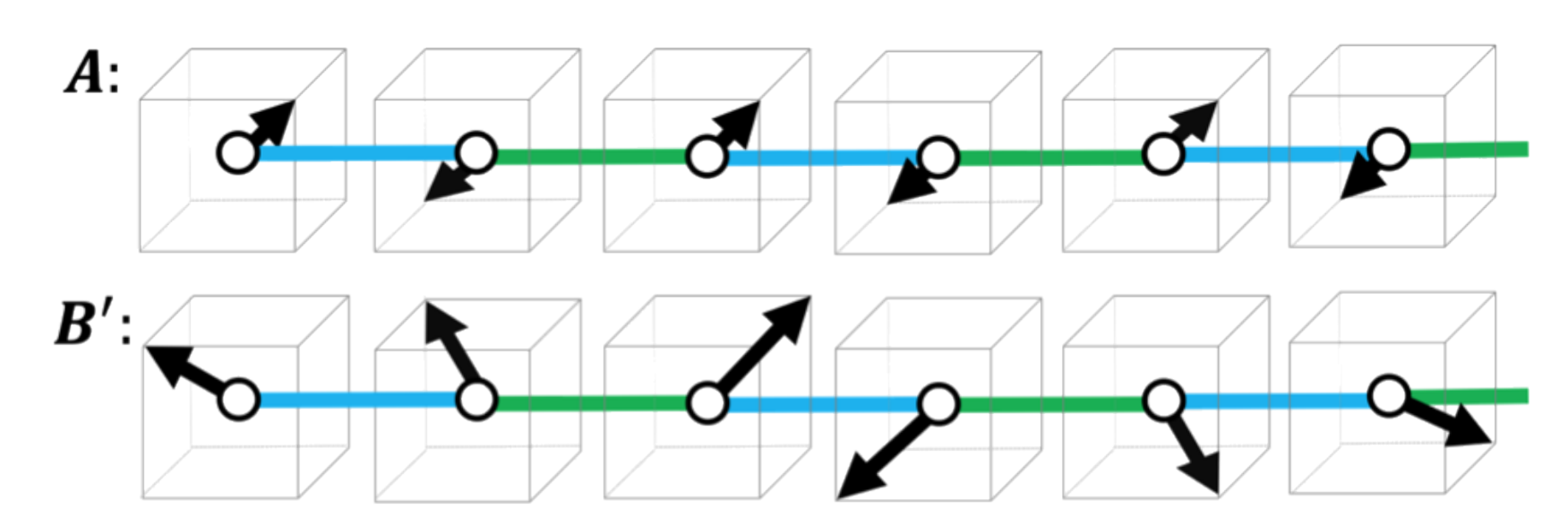}
\caption{Spin orientations in the original frame for two  representative  states in Fig. \ref{fig:spin_orders} corresponding to the $(1,1,1)$ and  $(1,-1,1)$ vertices, respectively.
Pattern $A$ is N\'eel, and pattern $B^\prime$ has a six-site periodicity. 
} 
\label{fig:spin_orient_orig}
\end{figure}

In the six-sublattice rotated frame, the spin orientations in the eight-fold degenerate ground states with an $O_h\rightarrow D_3$ symmetry breaking are \cite{Yang2020}
\bea
\vec{S}_{1+3n}&=&(\eta_1 a,\eta_2 a,\eta_3 b)^T,\nn\\
\vec{S}_{2+3n}&=&(\eta_1 a,\eta_2 b,\eta_3 a)^T,\nn\\
\vec{S}_{3+3n}&=&(\eta_1 b,\eta_2 a,\eta_3 a)^T,
\label{eq:spin_orig_OhD3}
\eea
in which $\eta_j=\pm1$ ($j=1,2,3$), and $a,b$ are real numbers (in principle not necessarily positive).
The ``center of mass" directions of the three spins within a unit cell  are plotted as the eight blue circles in Fig. \ref{fig:spin_orders}.
Here we want to emphasize that
although the spin orientations within the eight degenerate ground states are all translationally invariant by three sites, they exhibit different patterns in the original frame.
For the two light blue circles in Fig. \ref{fig:spin_orders}, 
the two corresponding  ground states exhibit a N\'eel order in the original frame;
on the other hand, the other six dark blue circles exhibit a six-site periodicity in the original frame.
More explicitly, in Fig. \ref{fig:spin_orient_orig}, the spin orientations in the ground states corresponding to $(1,1,1)$ and  $(1,-1,1)$ vertices are plotted in the original frame which have N\'eel and six-site periodic patterns, respectively.

For later convenience,  here we discuss the invariant correlation functions for the $O_h\rightarrow D_3$ order in the six-sublattice rotated frame.
In general, the ground state of a finite size system calculated in DMRG numerics may be an arbitrary linear combination of the several nearly degenerate  states (becoming exactly degenerate only in the thermodynamic limit)  with the coefficients depending on numerical details.
Thus the numerical results may  not represent the true values of the correlation functions in the thermodynamic limit. 
In addition, when performing the finite size scaling, such arbitrariness  may lead to a random oscillation of the correlation functions by varying the system size which does not exhibit the correct finite size scaling behavior.
Therefore, one needs to construct invariant correlation functions which take the same values in the ground state subspace.
As an example, consider an off-diagonal correlation function $\langle S_1^x S_{1+3n}^y\rangle$. 
It equals $a^2$ for the ground state corresponding to the $(1,1,1)$ vertex in Fig. \ref{fig:spin_orders}.
On the other hand, for the state corresponding to the $(1,-1,1)$ vertex, the value is $-a^2$.
In a finite size system, it is very plausible that the finite size ground state has components on both states at $(1,\pm1,1)$-vertices. 
Then there will be a cancellation so that  $\langle S_1^x S_{1+3n}^y\rangle$ acquires some arbitrary value.
This example illustrates the importance of constructing invariant correlation functions. 
 
In fact, for the $O_h\rightarrow D_3$ symmetry breaking, it is straightforward to see that all the diagonal correlation functions 
\bea
\left<S_i^\alpha S_{j+3n}^\alpha\right>
\eea
 are invariant correlation functions,
 where  $\alpha=x,y,z$, and  $i,j=1,2,3$.

\subsection{The spin-nematic order}
\label{subsec:sn}

We emphasize that the true magnetic order in the region $\phi\in[\phi_c^\prime,\phi_c]$ is the classical $O_h\rightarrow D_3$ order.
However,
at short and intermediate distances, 
 the system behaves as having a spin-nematic order with the symmetry breaking pattern $O_h\rightarrow D_{3d}$.
Since $|O_h/D_{3d}|=4$, the degeneracy is four.

The spin-nematic order parameters can be determined from the $O_h\rightarrow D_{3d}$ symmetry breaking pattern. 
It turns out that there are four independent spin-nematic order parameters.
In one of the four degenerate ground states, 
they are 
\bea
c&=&\bra{\Omega_e} S_1^xS_2^z \ket{\Omega_e}=\bra{\Omega_e} S_1^yS_2^x \ket{\Omega_e}=\bra{\Omega_e} S_2^xS_3^z \ket{\Omega_e}\nn\\
&=&\bra{\Omega_e} S_2^zS_3^y\ket{\Omega_e}
=\bra{\Omega_e} S_3^y S_4^x\ket{\Omega_e}
=\bra{\Omega_e} S_3^zS_4^y\ket{\Omega_e},\nn\\
\label{eq:quadrupole_order_c}
d&=&\bra{\Omega_e} S_1^xS_2^y \ket{\Omega_e}=\bra{\Omega_e} S_1^zS_2^x \ket{\Omega_e}=\bra{\Omega_e} S_2^yS_3^z \ket{\Omega_e}\nn\\
&=&\bra{\Omega_e} S_2^zS_3^x\ket{\Omega_e}
=\bra{\Omega_e} S_3^x S_4^y\ket{\Omega_e}
=\bra{\Omega_e} S_3^y S_4^z\ket{\Omega_e},\nn\\
\label{eq:quadrupole_order_d}
e&=&\bra{\Omega_e} S_1^yS_2^z \ket{\Omega_e}=\bra{\Omega_e} S_2^x S_3^y\ket{\Omega_e}=\bra{\Omega_e} S_3^zS_4^x \ket{\Omega_e},\nn\\
\label{eq:quadrupole_order_e}
f&=&\bra{\Omega_e} S_1^zS_2^y \ket{\Omega_e}=\bra{\Omega_e} S_2^y S_3^x\ket{\Omega_e}=\bra{\Omega_e} S_3^xS_4^z \ket{\Omega_e}.\nn\\
\label{eq:quadrupole_order_f}
\eea
Detailed derivations of Eq. (\ref{eq:quadrupole_order_f}) are included in Appendix \ref{app:spin_nematic}.

The other three ground states can be obtained from $\ket{\Omega_\alpha}=R(\hat{\alpha},\pi)\ket{\Omega_e}$ ($\alpha=x,y,z$),
since $R(\hat{\alpha},\pi)$ are representative operations in the equivalent classes in $O_h/D_{3d}$.
As an example, the expectation values of $e$ in the states $\ket{\Omega_\alpha}$ ($\alpha=x,y,z$) are given by
\bea
e&=& \bra{\Omega_x} S_1^y S_2^z \ket{\Omega_x} = -\bra{\Omega_x} S_3^z S_4^x \ket{\Omega_x} =-\bra{\Omega_x} S_2^x S_3^y \ket{\Omega_x},\nn\\
&=& -\bra{\Omega_y} S_1^y S_2^z \ket{\Omega_y} = \bra{\Omega_y} S_3^z S_4^x \ket{\Omega_y} =-\bra{\Omega_y} S_2^x S_3^y \ket{\Omega_y},\nn\\
&=&- \bra{\Omega_z} S_1^y S_2^z \ket{\Omega_z} = -\bra{\Omega_z} S_3^z S_4^x \ket{\Omega_z} =\bra{\Omega_z} S_2^x S_3^y \ket{\Omega_z}.\nn\\
\label{eq:e_xyz}
\eea

\subsection{The $O_h\rightarrow D_4$ order}

In this subsection, we briefly summarize the $O_h\rightarrow D_4$ phase in Fig. \ref{fig:phase_diagram} which has been analyzed in Ref. \onlinecite{Yang2019}.

\begin{figure}
\center
\includegraphics[width=7.5cm]{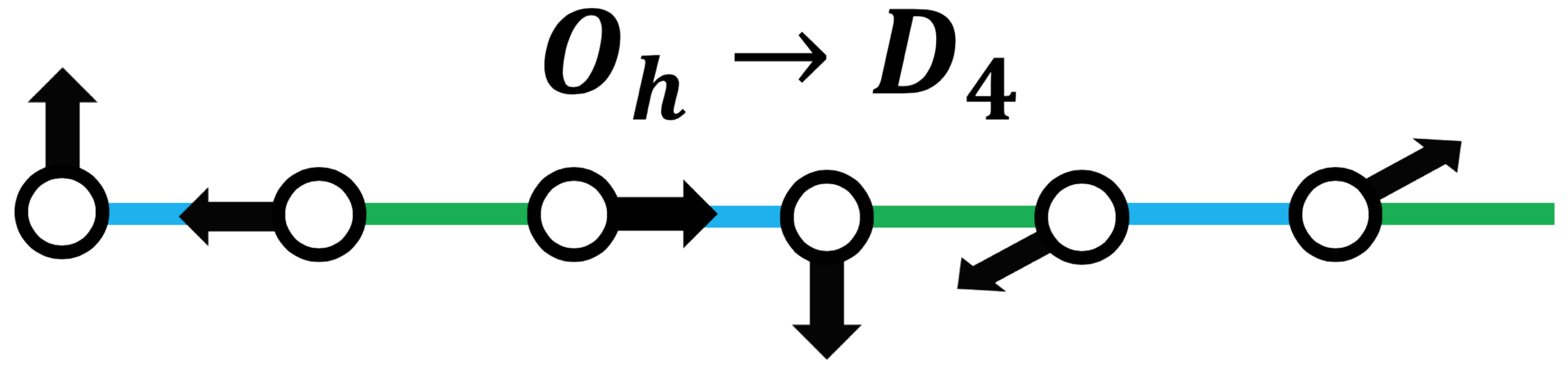}
\caption{Spin orientations in the original frame for the $O_h\rightarrow D_4$ order.
} 
\label{fig:spin_orient_orig_OhD4}
\end{figure}

In the six-sublattice rotated frame, the spin orientations in the six-fold degenerate ground states are \cite{Yang2019}
\bea
\vec{S}_{1+3n}=\eta b^\prime\hat{\alpha},~\vec{S}_{2+3n}=\eta a^\prime\hat{\alpha},~\vec{S}_{3+3n}=\eta a^\prime\hat{\alpha},
\label{eq:spin_orig_OhD4}
\eea
in which $\eta=\pm1$, and $\alpha=x,y,z$. 
In the original frame, they exhibit six-site periodicities. 
For example, the spin orientations for the state of $\alpha=z$ in the original frame are 
\begin{flalign}
&\vec{S}_{1+6n}=\hat{z},~
\vec{S}_{2+6n}=-\hat{y},~
\vec{S}_{3+6n}=\hat{y},\nn\\
&\vec{S}_{4+6n}=-\hat{z},~
\vec{S}_{5+6n}=\hat{x},~
\vec{S}_{6+6n}=-\hat{x}.
\label{eq:spin_orient_orig_OhD4}
\end{flalign}
A plot of Eq. (\ref{eq:spin_orient_orig_OhD4}) is shown in Fig. \ref{fig:spin_orient_orig_OhD4}.

We also discuss the invariant correlation functions for the $O_h\rightarrow D_4$ order in the six-sublattice rotated frame,
which are needed in DMRG calculations.  
There are in total ten invariant correlation functions (for detailed discussions, see Appendix \ref{sec:inv_corr}).
For our purposes, we only need the following two:
\bea
a^{\prime 2}&=&\langle S^x_1S^x_{2+3n}+S^y_1S^y_{3+3n}+S^z_2S^z_{3+3n}\rangle,\nn\\
0&=&\langle S_1^yS_{1+3n}^z+S_1^xS_{3+3n}^z+S_2^xS_{3+3n}^y\rangle,
\label{eq:OhD4_off}
\eea
in which $\langle ... \rangle$ denotes the expectation value.
In particular, since the average takes the same value within the ground state subspace, we do not need to specify which ground state the expectation value is taken in Eq. (\ref{eq:OhD4_off}).

\section{Misleading spin-nematic order at short and intermediate length scales}
\label{sec:spin_nematic}

In this section, we show that at short and intermediate length scales, the system behaves as having a spin-nematic order.
In particular, the time reversal symmetry remains preserved. 
In later sections, we will demonstrate that the time reversal symmetry is further broken at long distances, and the system transits into the classical $O_h\rightarrow D_3$ order.
Therefore, there is a two-step symmetry breaking $O_h\rightarrow D_{3d}\rightarrow D_3$
as the length scale is increased. 

\subsection{Ground state degeneracy}

\begin{figure}[h]
\includegraphics[width=7.5cm]{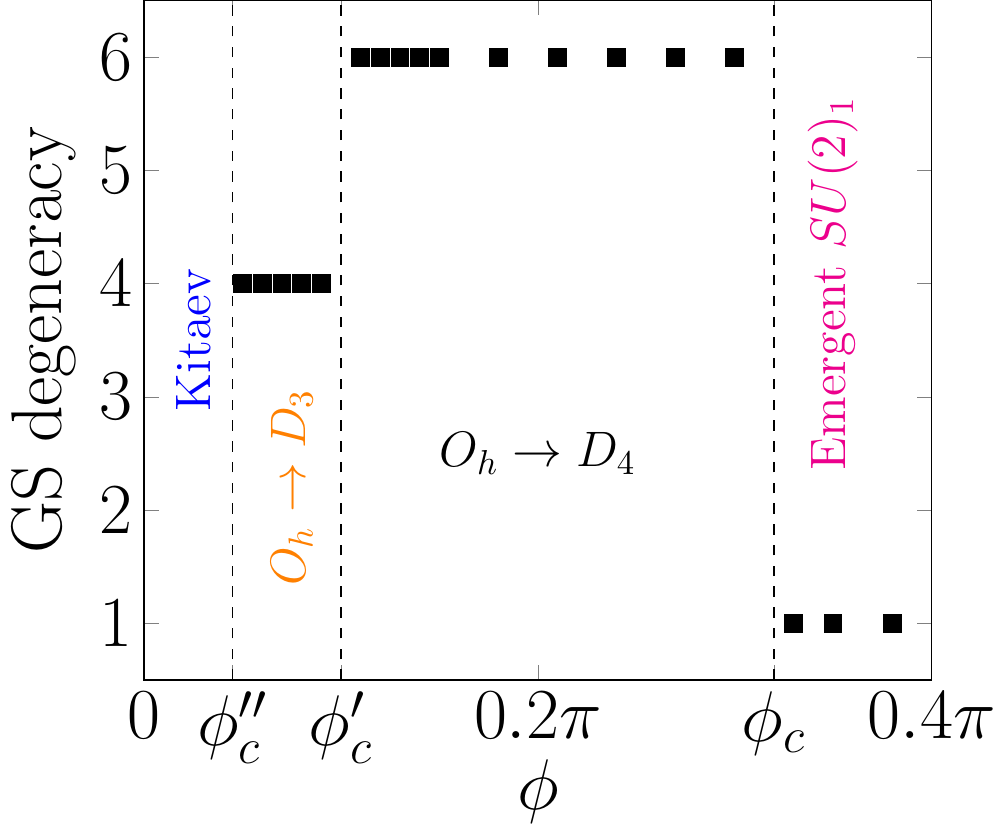}
\caption{Ground state degeneracy as a function of $\phi$. 
The three phase transition points are $\phi_c\simeq 0.33\pi$, $\phi_c^\prime\simeq 0.10\pi$, $\phi_c^{\prime\prime}\simeq 0.05\pi$. 
ED numerics are performed on a system of $L=24$ sites with periodic boundary conditions.
} \label{fig:quadru_degen}
\end{figure}

\begin{table}
                \begin{tabular}[t]{|c|c|c|c|}
		  \multicolumn{4}{c}{}\\ \hline
		  &  $\phi=0.06\pi$ & $\phi=0.07\pi$ & $\phi=0.08\pi$\\ \hline
$E_1$ &\tikzmark{top left 1a}-3.822078&\tikzmark{top left 1b}-3.821406&\tikzmark{top left 1c}-3.822237\\
$E_2-E_1$ &  $7.2\cdot 10^{-5}$&$1.0\cdot 10^{-4}$ & $1.7\cdot 10^{-4}$\\
$E_3-E_1$ &  $7.2\cdot 10^{-5}$&$1.0\cdot 10^{-4}$ & $1.7\cdot 10^{-4}$\\
$E_4-E_1$ &  $7.2\cdot 10^{-5}$\tikzmark{bottom right 1a} & $1.0\cdot 10^{-4}$\tikzmark{bottom right 1b} & $1.7\cdot 10^{-4}$\tikzmark{bottom right 1c}\\
$E_5-E_1$ &  $7.7\cdot 10^{-4}$ & $1.27\cdot 10^{-3}$ & $1.84\cdot 10^{-3}$\\
$E_6-E_1$ &  $7.7\cdot 10^{-4}$ & $1.27\cdot 10^{-3}$ & $1.84\cdot 10^{-3}$\\
$E_7-E_1$ &  $8.6\cdot 10^{-4}$ & $1.51\cdot 10^{-3}$ & $2.09\cdot 10^{-3}$\\ \hline
                \end{tabular}
			 	\DrawBox[ultra thick, red]{top left 1a}{bottom right 1a}
                \DrawBox[ultra thick, red]{top left 1b}{bottom right 1b}
                \DrawBox[ultra thick, red]{top left 1c}{bottom right 1c}
                \hfill
                \begin{tabular}[t]{|c|c|c|}
		  \multicolumn{3}{c}{}\\ \hline
		  & $\phi=0.09\pi$ & $\phi=0.10\pi$ \\ \hline
$E_1$&\tikzmark{top left 1d}     -3.825134&\tikzmark{top left 1e}     -3.830593\\
$E_2-E_1$&$3.35\cdot 10^{-4}$&$6.32\cdot 10^{-4}$\\
$E_3-E_1$&$3.35\cdot 10^{-4}$&$6.32\cdot 10^{-4}$\\
$E_4-E_1$&$3.35\cdot 10^{-4}$\tikzmark{bottom right 1d}&$6.32\cdot 10^{-4}$\tikzmark{bottom right 1e}\\
$E_5-E_1$&$2.28\cdot 10^{-3}$&$2.044\cdot 10^{-3}$\\
$E_6-E_1$&$2.28\cdot 10^{-3}$&$2.044\cdot 10^{-3}$\\
$E_7-E_1$&$2.33\cdot 10^{-3}$&$2.044\cdot 10^{-3}$\\ \hline
                \end{tabular}
                \DrawBox[ultra thick, red]{top left 1d}{bottom right 1d}
                \DrawBox[ultra thick, red]{top left 1e}{bottom right 1e}
				\hfill
\caption{Energies of several lowest lying states computed with
Lanczos Exact Diagonalization. The data refer to $L=24$ sites.
The four energies enclosed by the red squares are approximately degenerate at the corresponding $\phi$'s.
}
\label{table:energies_QP}
\end{table}

In Fig. \ref{fig:quadru_degen}, the ground state degeneracy is shown as a function of $\phi$. 
ED calculations are performed on a system of $L=24$ sites with periodic boundary conditions.
Three phase transitions $\phi_c,\phi_c^\prime,\phi_c^{\prime\prime}$  can be identified based on the ground state degeneracy, where $\phi_c\simeq 0.33\pi$, $\phi_c^\prime\simeq 0.10\pi$, and $\phi_c^{\prime\prime}\simeq 0.05\pi$.
Here we note that iDMRG gives a shifted value of $\phi_c^{\prime\prime}$ as will be discussed in Section \ref{sec:Kitaev}.
When $\phi>\phi_c$, the ground state was nondegenerate which corresponds to the ``Emergent SU(2)$_1$ I" phase as shown in Fig. \ref{fig:phase_diagram}.
For $\phi\in[\phi_c^\prime,\phi_c]$, the ground states are six-fold degenerate, corresponding to the $O_h\rightarrow D_4$ phase which has been identified in Ref. \onlinecite{Yang2019}.
In the range $\phi_c^{\prime\prime}<\phi<\phi_c^\prime$,
numerics provide evidence for a four-fold ground state degeneracy,
implying a spin ordering different from $O_h\rightarrow D_4$.
The energies of the first seven states are displayed in Table \ref{table:energies_QP}, from which the four-fold ground state degeneracy can be observed.
Indeed, the four energies enclosed by the red square at each angle $\phi$ are approximately degenerate, and they are separated from the other states with an energy gap much larger than the splitting among themselves.
Finally, for $\phi<\phi_c^{\prime\prime}$,
we were unable to find a definite value of the ground state degeneracy.
The degeneracy has strong finite size dependences,
and there is no clear energy separation between some low lying (presumably ground state) multiplet and the excited states.
Therefore, no value of degeneracy is assigned to the region of $\phi<\phi_c^{\prime\prime}$, which is denoted as the ``Kitaev" phase in Fig. \ref{fig:quadru_degen}.

Clearly, the four-fold ground state degeneracy in the region $[\phi_c^{\prime\prime},\phi_c^\prime]$ is not consistent with the classical $O_h\rightarrow D_3$ order which is eight-fold degenerate.
In fact, in Appendix \ref{app:inconsistency}, we are able to prove that this four-fold degeneracy is not consistent with any rank-1 magnetic order, provided that the translational symmetry $T_{3a}$ is not broken. 
Then the simplest possibility is the rank-2 spin-nematic order.

\subsection{Numerical evidence for spin-nematic order for $L=24$ sites}

In this subsection, we provide numerical evidence for the spin-nematic order on a system of  $L=24$ sites.

Define a spin-nematic-$\lambda$ field $h_{\lambda}$ as $-h_{\lambda} L \hat{Q}_{\lambda}$, where the spin-nematic operator $\hat{Q}_\lambda$ is  the sum of the order parameters in Eq. (\ref{eq:quadrupole_order_f}),
where $\lambda=c,d,e,f$.
For example, $\hat{Q}_e$ is given by
\begin{flalign}
\hat{Q}_e=\frac{1}{L}\sum_{n}( S_{1+3n}^y S_{2+3n}^z+S_{2+3n}^x S_{3+3n}^y+S_{3+3n}^z S_{4+3n}^x),
\label{eq:Qe}
\end{flalign}
and the other three spin-nematic orders can be obtained similarly.
Consider a small $h_{\lambda}$ field which satisfies
$\Delta E\ll  |e||h_{\lambda}|L \ll E_g$, in which $L$ is the system size, $E_g$ is the excitation gap, and $\Delta E$ is the finite size splitting of the ground state quartet at zero field.  
Suppose the system has spin-nematic orders defined in Sec. \ref{subsec:sn},
then such choice of $h_{\lambda}$ gives rise to a degenerate perturbation within the four-dimensional ground state subspace, and at the same time, no mixing between the ground states and the excited states is induced. 
According to Eqs. (\ref{eq:quadrupole_order_e},\ref{eq:e_xyz}),  the energies of the four ground states under $h_{\lambda}$ are
\bea
\delta E(\ket{\Omega_e})&=&-\lambda h_{\lambda}L,\nn\\
\delta E(\ket{\Omega_\alpha})&=&\frac{1}{3}\lambda h_{\lambda} L,
\label{eq:deltaE}
\eea
in which $\alpha=x,y,z$, and the energy $\delta E$ is measured with respect to the zero field case.
Therefore, if $\text{sgn}(h_{\lambda})=\text{sgn}(\lambda)$, 
$\ket{\Omega_e}$ is the ground state which is nondegenerate
with an energy lowered by an amount $|\lambda h_{\lambda}|L$.
On the other hand, if $\text{sgn}(h_\lambda)=-\text{sgn}(\lambda)$, 
the ground states are three-fold degenerate,
and in fact, $\ket{\Omega_x},\ket{\Omega_y},\ket{\Omega_z}$ have the same energy which is lower than the energy at zero field by an amount $\frac{1}{3}|\lambda h_\lambda|L$.
In this way, the sign of the spin-nematic order parameter can be obtained from that of the corresponding spin-nematic field by inspecting the change of the ground state degeneracy. 
In addition, let $\delta E_g (h_\lambda)$ be the ground state energy change with a field $h_\lambda$.
Then according to Eq. (\ref{eq:deltaE}), we obtain
\bea
\frac{\delta E_g(|h_\lambda|\text{sgn}(\lambda))}{\delta E_g(-|h_\lambda|\text{sgn}(\lambda))}=3.
\label{eq:energyratio_3}
\eea 
in which $\lambda=c,d,e,f$.

\begin{table}
                \begin{tabular}[t]{|c|c|c|c|}
		  \multicolumn{4}{c}{}\\ \hline
		 (a) & $h_{c}=10^{-3}$ & No field & $h_{c}=-10^{-3}$\\ \hline 
		$E_1$&\tikzmark{top left 1c}-3.8226887\tikzmark{bottom right 1c}
		&\tikzmark{top left 2c}-3.8220786&\tikzmark{top left 3c}-3.8222885\\ 
		$E_2-E_1$&$7.99\cdot10^{-4}$&$6.90\cdot10^{-5}$&$3.51\cdot10^{-5}$\\
		$E_3-E_1$&$8.58\cdot10^{-4}$&$6.90\cdot10^{-5}$&$3.52\cdot10^{-5}$\tikzmark{bottom right 3c}\\ 
		$E_4-E_1$&$8.58\cdot10^{-4}$&$6.90\cdot10^{-5}$\tikzmark{bottom right 2c}&$7.68\cdot10^{-4}$\\ 
		$E_5-E_1$&$1.14\cdot10^{-3}$&$7.68\cdot10^{-4}$&$7.68\cdot10^{-4}$\\ 
		$E_6-E_1$&$1.14\cdot10^{-3}$&$7.68\cdot10^{-4}$&$8.31\cdot10^{-4}$\\ \hline
                \end{tabular}
                \DrawBox[ultra thick, red]{top left 1c}{bottom right 1c}
                \DrawBox[ultra thick, blue]{top left 2c}{bottom right 2c}
                \DrawBox[ultra thick, green]{top left 3c}{bottom right 3c}
                \hfill
                \begin{tabular}[t]{|c|c|c|c|}
		  \multicolumn{4}{c}{}\\ \hline
		 (b) & $h_{d}=10^{-3}$ & No field & $h_{d}=-10^{-3}$\\ \hline 
		$E_1$&\tikzmark{top left 1d}-3.8225967\tikzmark{bottom right 1d}
		&\tikzmark{top left 2d}-3.8220786&\tikzmark{top left 3d}-3.8222538\\ 
		$E_2-E_1$&$6.945\cdot10^{-4}$&$6.90\cdot10^{-5}$&$3.60\cdot10^{-5}$\\
		$E_3-E_1$&$7.545\cdot10^{-4}$&$6.90\cdot10^{-5}$&$3.60\cdot10^{-5}$\tikzmark{bottom right 3d}\\ 
		$E_4-E_1$&$7.545\cdot10^{-4}$&$6.90\cdot10^{-5}$\tikzmark{bottom right 2d}&$7.75\cdot10^{-4}$\\ 
		$E_5-E_1$&$1.105\cdot10^{-3}$&$7.68\cdot10^{-4}$&$7.75\cdot10^{-4}$\\ 
		$E_6-E_1$&$1.105\cdot10^{-3}$&$7.68\cdot10^{-4}$&$8.40\cdot10^{-4}$\\ \hline
                \end{tabular}
                \DrawBox[ultra thick, red]{top left 1d}{bottom right 1d}
                \DrawBox[ultra thick, blue]{top left 2d}{bottom right 2d}
                \DrawBox[ultra thick, green]{top left 3d}{bottom right 3d}
                \hfill
                \begin{tabular}[t]{|c|c|c|c|}
		  \multicolumn{4}{c}{}\\ \hline
		 (c) & $h_{e}=10^{-3}$ & No field & $h_{e}=-10^{-3}$\\ \hline 
		$E_1$&\tikzmark{top left 1e}-3.8255339\tikzmark{bottom right 1e}
		&\tikzmark{top left 2e}-3.8220786&\tikzmark{top left 3e}-3.8234555\\ 
		$E_2-E_1$&$2.339\cdot10^{-3}$&$6.90\cdot10^{-5}$&$7.20\cdot10^{-6}$\\
		$E_3-E_1$&$2.339\cdot10^{-3}$&$6.90\cdot10^{-5}$&$7.30\cdot10^{-6}$\tikzmark{bottom right 3e}\\ 
		$E_4-E_1$&$2.495\cdot10^{-3}$&$6.90\cdot10^{-5}$\tikzmark{bottom right 2e}&$5.60\cdot10^{-4}$\\ 
		$E_5-E_1$&$2.493\cdot10^{-3}$&$7.68\cdot10^{-4}$&$5.60\cdot10^{-4}$\\ 
		$E_6-E_1$&$2.537\cdot10^{-3}$&$7.68\cdot10^{-4}$&$7.78\cdot10^{-4}$\\ \hline
                \end{tabular}
                \DrawBox[ultra thick, red]{top left 1e}{bottom right 1e}
                \DrawBox[ultra thick, blue]{top left 2e}{bottom right 2e}
                \DrawBox[ultra thick, green]{top left 3e}{bottom right 3e}
                \hfill                
                \begin{tabular}[t]{|c|c|c|c|}
		  \multicolumn{4}{c}{}\\ \hline
		 (d) & $h_{f}=10^{-3}$ & No field & $h_{f}=-10^{-3}$\\ \hline 
		$E_1$&\tikzmark{top left 1f}-3.8242832\tikzmark{bottom right 1f}
		&\tikzmark{top left 2f}-3.8220786&\tikzmark{top left 3f}-3.8228937\\ 
		$E_2-E_1$&$1.847\cdot10^{-3}$&$6.90\cdot10^{-5}$&$2.00\cdot10^{-6}$\\
		$E_3-E_1$&$1.847\cdot10^{-3}$&$6.90\cdot10^{-5}$&$2.00\cdot10^{-6}$\tikzmark{bottom right 3f}\\ 
		$E_4-E_1$&$1.194\cdot10^{-3}$&$6.90\cdot10^{-5}$\tikzmark{bottom right 2f}&$6.21\cdot10^{-4}$\\ 
		$E_5-E_1$&$1.194\cdot10^{-3}$&$7.68\cdot10^{-4}$&$6.21\cdot10^{-4}$\\ 
		$E_6-E_1$&$2.070\cdot10^{-3}$&$7.68\cdot10^{-4}$&$8.03\cdot10^{-4}$\\ \hline
                \end{tabular}
                \DrawBox[ultra thick, red]{top left 1f}{bottom right 1f}
                \DrawBox[ultra thick, blue]{top left 2f}{bottom right 2f}
                \DrawBox[ultra thick, green]{top left 3f}{bottom right 3f}                                                                                                
\caption{Energies of several lowest lying states computed with 
Lanczos Exact Diagonalization. The data refer to $L=24$ sites, and $\phi=0.06\pi$.}
\label{table:enegies}
\end{table}

In Table \ref{table:enegies}, the energies of the six lowest states are displayed under different spin-nematic fields $h_\alpha$ ($\alpha=c,d,e,f$) at $\phi=0.06\pi$.
The results are obtained from ED calculations on a system of $L=24$ sites with periodic boundary conditions.
As can be seen from Table \ref{table:enegies}, the four states circled by the blue lines are separated from the other two states by an energy $\simeq 6.9\times 10^{-4}$, which is one order of magnitude larger than the energy splitting among the four states which is $\simeq 0.77\times 10^{-4}$.
This provides numerical evidence for the four-fold ground state degeneracy at zero field as discussed in Fig. \ref{fig:quadru_degen}.
On the other hand, as shown in Table \ref{table:enegies}, the system is nondegenerate under positive spin-nematic fields for all the four $h_\alpha$'s where $\alpha=c,d,e,f$,
but approximately three-fold degenerate when the fields are negative.
According to the previous discussions, this provides numerical evidence for the spin-nematic order parameters $c,d,e,f$ to be all positive.
In addition, we check if the relations in Eq. (\ref{eq:energyratio_3}) for the energy changes are satisfied.
According to Table \ref{table:enegies}, the ratios $r=\delta E_g(|h_\lambda|\text{sgn}(\lambda))/\delta E_g(-|h_\lambda|\text{sgn}(\lambda))$ are
\bea
\begin{array}{ccccc}
\lambda & c&d&e&f\\
r & 2.91& 2.96& 2.51 & 2.70.
\end{array}
\label{eq:results_ratio_3}
\eea
As can be seen from Eq. (\ref{eq:results_ratio_3}), while the ratios for $\lambda=c,d$   agree well with $3$, there are slight deviations of $r$ from $3$ for $e,f$.
In fact,  the values of $e,f$ are much larger than $c,d$ (see Appendix \ref{app:strength_sn}).
Hence, a $10^{-3}$ field is too large for $e$ and $f$ in the sense that the conditions $\Delta E\ll  |\lambda||h_\lambda|L \ll E_g$ are spoiled when $\lambda=e,f$.
In these cases, $\delta E_g$ also involves the contributions from many excited states, not just the ground state quartet. 
Because of this reason, the relation in Eq. (\ref{eq:energyratio_3}) is not satisfied to an excellent level for $h_e,h_f\sim 10^{-3}$. 
A better agreement of $r$ with $3$ can be obtained for $e,f$ by choosing a much smaller value of the field $h_\lambda$.

The agreements with the predictions in Eq. (\ref{eq:energyratio_3}) provide strong evidence for the spin-nematic order for a system of $L=24$ sites. 
Notice in particular that the non-degenerate ground state under positive $h_\lambda$ ($\lambda=c,d,e,f$) fields imply the absence of rank-1 magnetic orders.
Otherwise the ground states has to be at least two-fold degenerate,
since the applied $h_\lambda$ fields preserves time reversal symmetry
and a sign change of the rank-1 orders by applying the time reversal operation will lead to a degenerate state with the same energy.

\section{Emergence of a rank-1 magnetic order at long distances}
\label{sec:emerge_rank1}

In this section, we provide numerical evidence for the emergence of a rank-1 magnetic order at long distances ($L\sim 100$). 
The method in Sec. \ref{sec:spin_nematic} cannot be applied, since the determination of ground state degeneracy is no longer available for large system sizes, which requires the knowledge of the energies of many finite size excited states. 
We will turn to an alternative method named as ``energy-field relation", which only involves the calculation of the (finite size) ground state energy, thereby can be easily pushed to much larger system sizes.

Suppose $\mathcal{O}$ is an order parameter.
By adding an $\mathcal{O}$-field, the Hamiltonian becomes
\bea
H=H_0-h\mathcal{O}.
\eea
Let $\ket{G_0}$ be the particular one of the ground states of $H_0$ which can be polarized by $\mathcal{O}$.
If the system has the $\mathcal{O}$-order, then $\bra{G_0}\mathcal{O}\ket{G_0}\neq 0$.
For small $h$, the energy change can be obtained from first order perturbation theory, i.e.,
\bea
\Delta E=E(h)-E(h=0)=-h\bra{G_0}\mathcal{O}\ket{G_0},
\label{eq:Delta_E}
\eea
which is linear in $h$.
On the other hand, If the system does not have $\mathcal{O}$-order,
the energy can only change by second order perturbation theory,
therefore $\Delta E\propto h^2$.
Hence, by looking at the energy-field relation to check if it is linear or quadratic,  we are able to test possible order parameters. 

\begin{figure}
\center
\includegraphics[width=8.3cm]{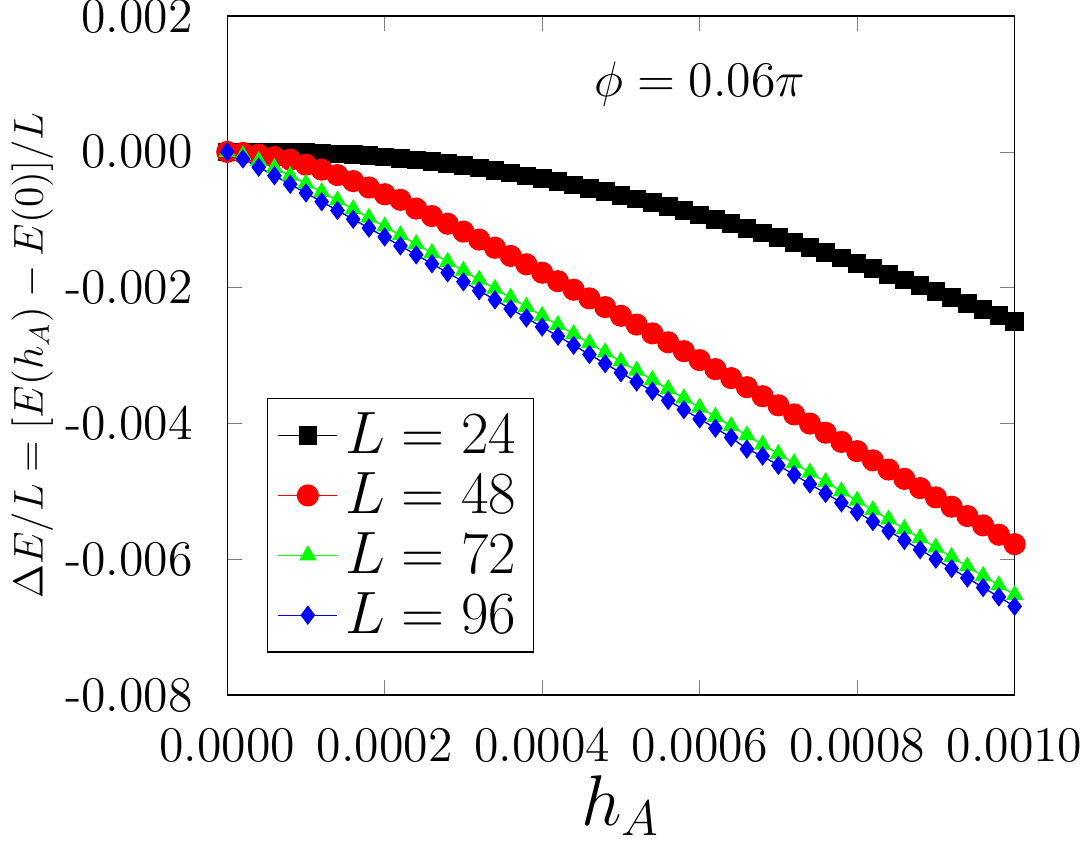}
\caption{Energy-field relations for  $h_{A}$ defined in Eq. (\ref{eq:H_hA}) for
$L=24$ (black curve), 
 $L=48$ (red curve), $L=72$ (green curve) and $L=96$ (blue curve).
Periodic boundary conditions are taken in the DMRG calculations.
} 
\label{fig:energy_field_hn}
\end{figure}

Let's consider a magnetic field $h_{A}$ along the $(1,1,1)$-direction. 
The Hamiltonian becomes
\bea
H=H^\prime-h_{A}\sum_n\frac{1}{\sqrt{3}}(S_n^x+S_n^y+S_n^z),
\label{eq:H_hA}
\eea
in which $H^\prime$ is defined in Eq. (\ref{eq:6rotated}).
The numerically calculated energy-field relation are displayed in Fig. \ref{fig:energy_field_hn} for systems of $L=24,48,72,96$ sites with periodic boundary conditions.
As can be observed from Fig. \ref{fig:energy_field_hn}, the energy-field relation is quadratic for $L=24$, indicating an absence of rank-1 orders consistent with the numerical results in Sec. \ref{sec:spin_nematic}.
However, the energy-field relation gets increasingly more linear as the system size is increased.
At $L=96$, the relation has already become very linear, implying the emergence of a rank-1 magnetic order. 
This shows that the spin-nematic order observed in Sec. \ref{sec:spin_nematic} is only an artifact at short and intermediate length scales. 

\section{Numerical evidence for the classical $O_h\rightarrow D_3$ order}

In this section, we provide enough numerical evidence to show that the emerged rank-1 magnetic order at $L\sim 100$ sites in Sec. \ref{sec:emerge_rank1} is a classical $O_h\rightarrow D_3$ order.
Throughout this section, we work in the six-sublattice rotated frame unless otherwise stated. 

\subsection{Correlation functions}

We first study the correlation functions both at zero fields and small magnetic fields.
We fix the angle $\phi$ at $0.06\pi$, which lies within the $O_h\rightarrow D_3$ phase according to Fig. \ref{fig:phase_diagram}.

\subsubsection{Zero field}

\begin{figure}
\center
\includegraphics[width=8.5cm]{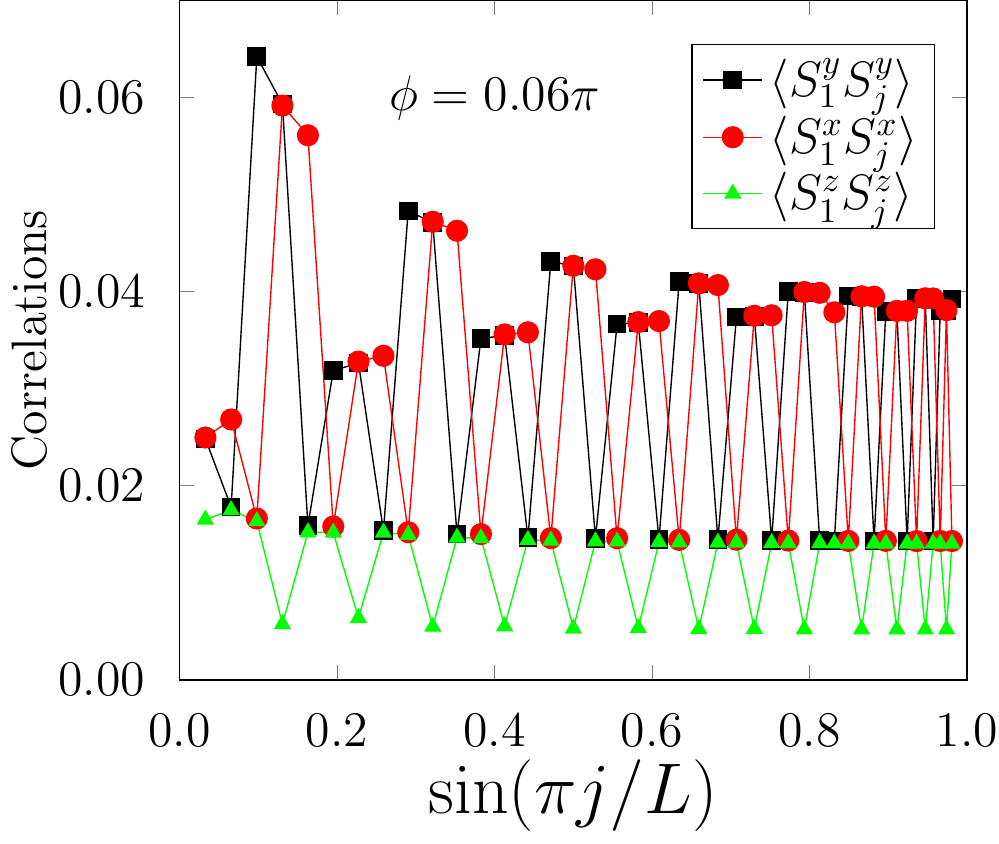}
\caption{Correlation functions $\langle S_1^xS_j^x\rangle$ (red curve), $\langle S_1^yS_j^y\rangle$ (black) and $\langle S_1^zS_j^z\rangle$ (green)  as a function of $j$.
The calculations are carried out at zero field, and the system size is taken as $L=96$.
} 
\label{fig:hp_NOfield_OhD3_Corr_2}
\end{figure}

Fig. \ref{fig:hp_NOfield_OhD3_Corr_2} plots the diagonal correlation functions $\langle S_1^\alpha S_j^\alpha\rangle$ ($\alpha=x,y,z$) as a function of $j$ calculated without magnetic field on a system of $L=96$ sites with periodic boundary conditions, where $\phi=0.06\pi$.
Recall from Sec. \ref{sec:OhD3} that the diagonal correlation functions are invariant correlation functions for the $O_h\rightarrow D_3$ order.
Therefore, it is legitimate to calculate $\langle S_1^\alpha S_j^\alpha\rangle$.

According to Eq. (\ref{eq:spin_orig_OhD3}), 
at long distances, $\langle S_1^\alpha S_j^\alpha\rangle$ should be equal to
\bea
&\langle S_1^x S_{1+3n}^x\rangle = a^2,~ \langle S_1^x S_{2+3n}^x\rangle= a^2,~\langle S_1^x S_{3+3n}^x\rangle= ab,\nn\\
&\langle S_1^y S_{1+3n}^y\rangle = a^2,~ \langle S_1^y S_{2+3n}^y\rangle= ab,~\langle S_1^y S_{3+3n}^y\rangle= a^2,\nn\\
&\langle S_1^zS_{1+3n}^z\rangle = b^2,~ \langle S_1^z S_{2+3n}^z\rangle= ab,~\langle S_1^z S_{3+3n}^z\rangle= ab.\nn\\
\eea
As can be seen from Fig. \ref{fig:hp_NOfield_OhD3_Corr_2}, these patterns are indeed satisfied. 
In particular, the values of $a^2,ab,b^2$ can be read as
\bea
a^2\sim 0.04,  ~ab\sim 0.014,~b^2\sim 0.005.
\label{eq:a2b2ab}
\eea
Notice that there is the relation $a^2\cdot b^2=(ab)^2$.
Indeed, $0.04\times 0.005=2\times 10^{-4}$ is pretty close to $0.014^2=1.96\times 10^{-4}$. 
This provides evidence for the $O_h\rightarrow D_3$ order.

However, we note that the evidence is not sufficient.
Denote $\ket{\eta \hat{\alpha}}$ ($\eta=\pm1$, $\alpha=x,y,z$) as the six degenerate ground states with $O_h\rightarrow D_4$ order in which the ``center of mass" direction of the three spins within a unit cell is pointing along the $\eta\hat{\alpha}$-direction as shown in Eq. (\ref{eq:spin_orig_OhD4}).
Then it can be easily checked that the state $\frac{1}{\sqrt{3}}(\ket{\hat{x}}+\ket{\hat{y}}+\ket{\hat{z}})$ would produce the same patterns of correlation functions as those in Fig. \ref{fig:hp_NOfield_OhD3_Corr_2}.
In the next two subsections, we test the correlation functions with small applied magnetic fields, which are able to resolve such potential concerns.

\subsubsection{Field along $(0,0,1)$-direction}

Suppose we apply a small field along the $(0,0,1)$-direction, and compute the diagonal correlation functions $\left<S_i^\alpha S_{j+3n}^\alpha\right>$ ($\alpha=x,y,z$).
If the symmetry breaking is $O_h\rightarrow D_4$, then $\left<S_i^x S_{j+3n}^x\right>=\left<S_i^y S_{j+3n}^y\right>=0$ but $\left<S_i^z S_{j+3n}^z\right>$ is nonzero,
since the state $\ket{\hat{z}}$ is now picked out as the ground state which has an energy lower  than the other five ground states at zero field. 
On the other hand, if the symmetry breaking is $O_h\rightarrow D_3$ order, then all of the three diagonal correlation functions are nonzero.

\begin{figure*}
\center
\includegraphics[width=14cm]{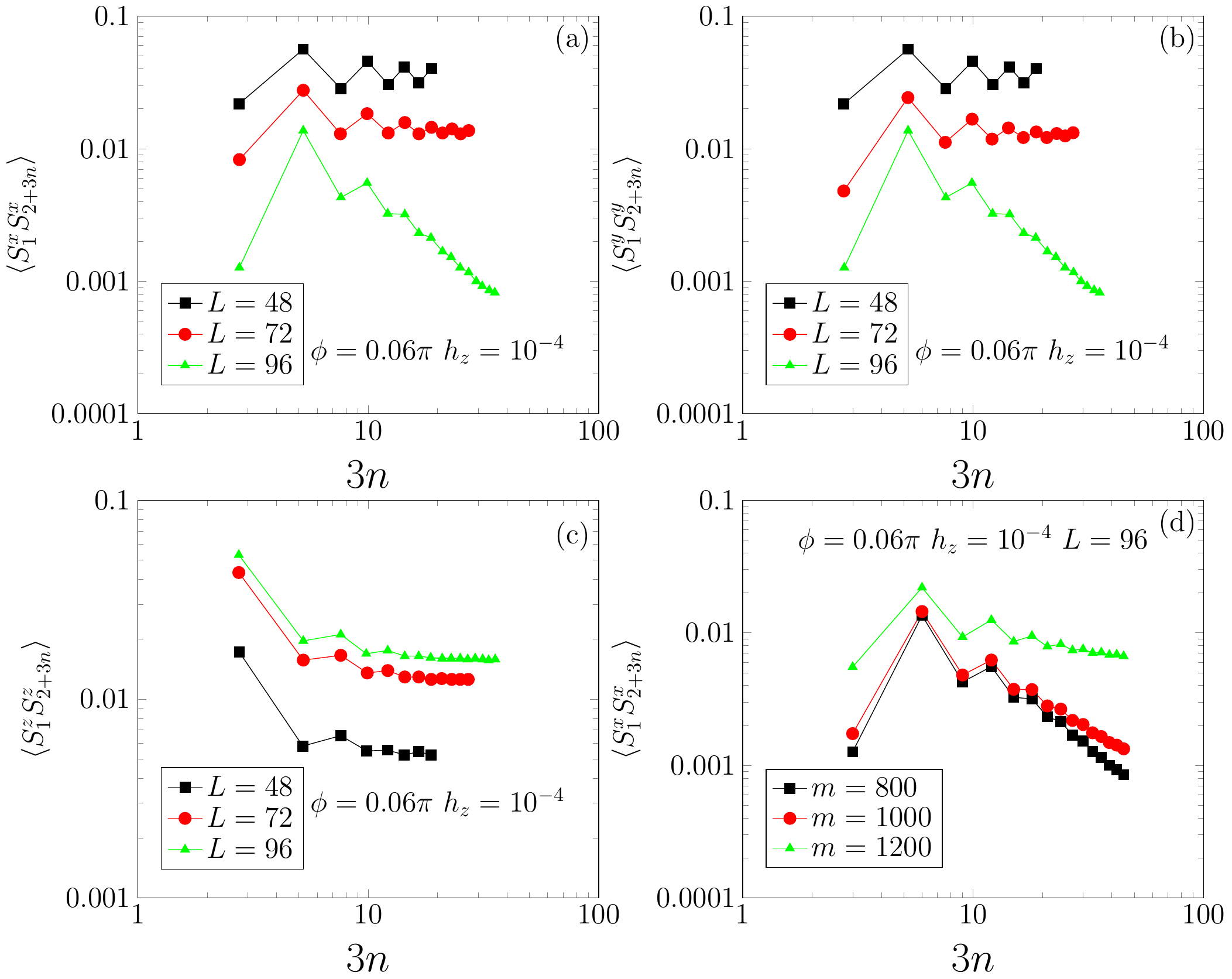}
\caption{Diagonal correlation functions (a) $\left<S_i^x S_{j+3n}^x\right>$, (b) $\left<S_i^y S_{j+3n}^y\right>$, (c) $\left<S_i^z S_{j+3n}^z\right>$,
and (d) $\left<S_i^x S_{j+3n}^x\right>$ by increasing DMRG accuracy,
where a $h_z=10^{-4}$ field along the $(0,0,1)$-direction is applied at $\phi=0.06\pi$.
DMRG numerical calculations are performed on $L=48,72,96$ sites.
In (d), the number of states $m$ kept in DMRG calculations is changed from $m=800$ to $m=1200$.
} 
\label{fig:corr_diagonal}
\end{figure*}

Fig. \ref{fig:corr_diagonal} (a,b,c) show the DMRG numerical results for the diagonal correlation functions $\left<S_i^\alpha S_{j+3n}^\alpha\right>$ ($\alpha=x,y,z$) at three system sizes $L=48,72,96$.
The field is chosen as $10^{-4}$ along the $(0,0,1)$-direction, and the angle is taken as $\phi=0.06\pi$.
As can be seen from Fig. \ref{fig:corr_diagonal} (a,b), the $\left<S_i^x S_{j+3n}^x\right>$ and $\left<S_i^y S_{j+3n}^y\right>$ correlation functions are nonzero, indicating an $O_h\rightarrow D_3$ rather than $O_h\rightarrow D_4$ order.

However, potential concerns arise for $\left<S_i^x S_{j+3n}^x\right>$ and $\left<S_i^y S_{j+3n}^y\right>$.
In fact, according to Fig. \ref{fig:corr_diagonal} (a,b), the $L=96$ results seem to exhibit a long-distance decay behavior,
and it is not clear if these values go to zero at extremely long distances.
To clarify this issue, we have increased the DMRG accuracy  by increasing the number of states $m$ kept in DMRG calculations.
The numerical results by changing $m$ are displayed in Fig. \ref{fig:corr_diagonal} (d) for $\left<S_i^x S_{j+3n}^x\right>$ at $L=96$. 
It is clear that the tail is significantly raised up when $m$ reaches $1200$.
Therefore, we conclude that the decay behavior observed in Fig. \ref{fig:corr_diagonal} (a,b) at $L=96$ is possibly a numerical artifact. 

\subsubsection{Field along $(1,1,1)$-direction}
\label{subsubsec:field111}

\begin{figure}
\center
\includegraphics[width=8.3cm]{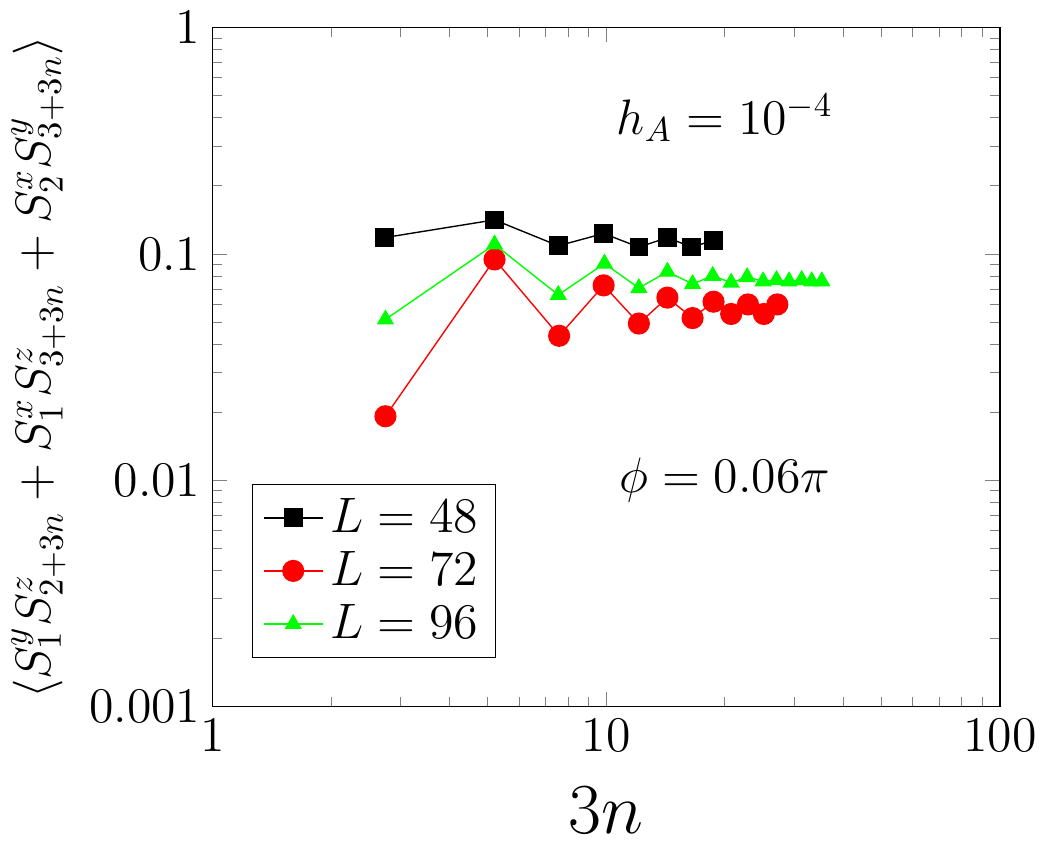}
\caption{$O_h\rightarrow D_4$ off-diagonal correlation function $\langle S_1^yS_{2+3n}^z+S_1^xS_{3+3n}^z+S_2^xS_{3+3n}^y\rangle$ with a $h_A=10^{-4}$ field along $(1,1,1)$-direction at $\phi=0.06\pi$.
} 
\label{fig:corr_offdiagonal_scaling}
\end{figure}

According to Eq. (\ref{eq:OhD4_off}), $\langle S_1^yS_{2+3n}^z+S_1^xS_{3+3n}^z+S_2^xS_{3+3n}^y\rangle$ is an $O_h\rightarrow D_4$ invariant correlation function which should be zero if the symmetry breaking is $O_h\rightarrow D_4$.
In Fig. \ref{fig:corr_offdiagonal_scaling}, we apply a $h_A=10^{-4}$ field along the $(1,1,1)$-direction at $\phi=0.06\pi$.
As can be seen from Fig. \ref{fig:corr_offdiagonal_scaling}, the correlation function $\langle S_1^yS_{2+3n}^z+S_1^xS_{3+3n}^z+S_2^xS_{3+3n}^y\rangle$ is nonvanishing, which invalidates the $O_h\rightarrow D_4$ order.
On the other hand, it is consistent with the $O_h\rightarrow D_3$ order, since for the ground state corresponding to the $(1,1,1)$-vertex in Fig. \ref{fig:spin_orders} (which is selected out by the small $(1,1,1)$-field), the off-diagonal correlation functions do not vanish. 

\subsubsection{Range of the $O_h\rightarrow D_3$ phase}

\begin{figure}
\center
\includegraphics[width=8.3cm]{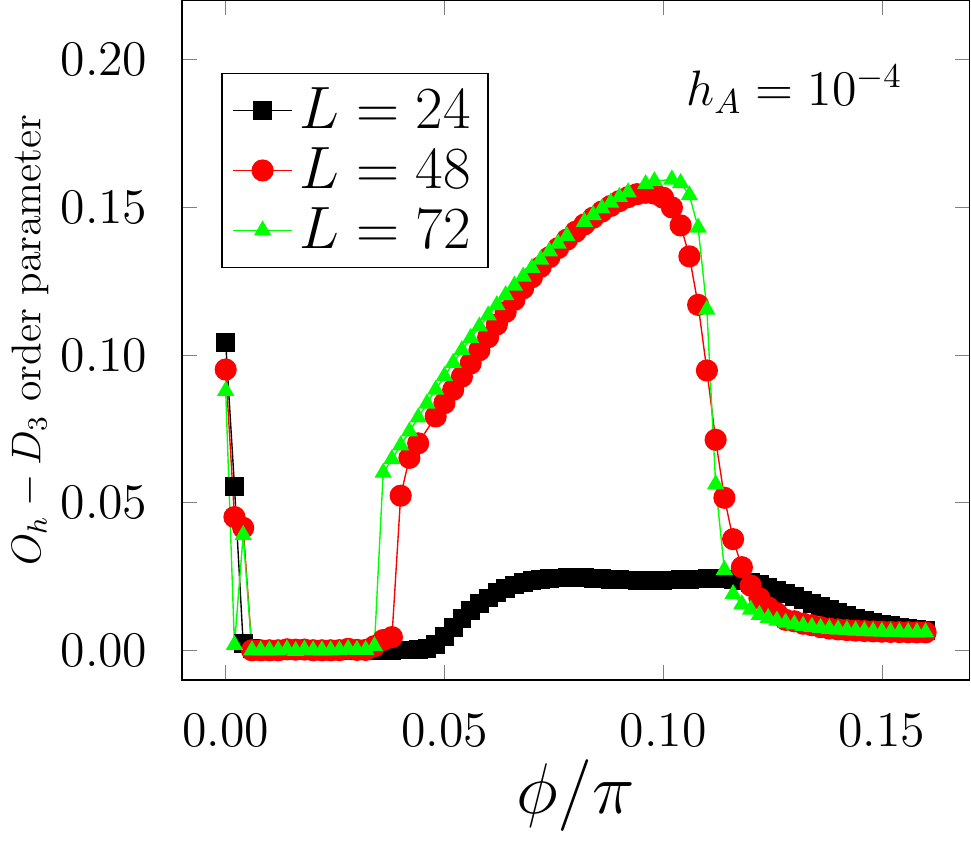}
\caption{Off-diagonal correlation function $\langle S_1^yS_{2+3n}^z+S_1^xS_{3+3n}^z+S_2^xS_{3+3n}^y\rangle$ as a function of $\phi$ in the presence of a small $h_A=10^{-4}$ field along $(111)$-direction.
} 
\label{fig:range_OhD3}
\end{figure}

We note that the test in Sec. \ref{subsubsec:field111} can be efficiently used to determine the range of the $O_h\rightarrow D_3$ phase. 

We  have calculated the off-diagonal $O_h\rightarrow D_4$ invariant correlation function $\langle S_1^yS_{2+3n}^z+S_1^xS_{3+3n}^z+S_2^xS_{3+3n}^y\rangle$ in the presence of a small $h_A=10^{-4}$ field along the $(1,1,1)$-direction.
In the $O_h\rightarrow D_3$ phase, since the ground state at the $(1,1,1)$-vertex is picked out, this correlation function is nonzero.
On the other hand, in the $O_h\rightarrow D_4$ phase, it vanishes due to the off-diagonal nature.

The DMRG numerical results are shown in Fig. \ref{fig:range_OhD3}.
As can be clearly seen, there is a sharp phase transition at $\phi\sim 0.035\pi$, which is $\phi_c^{\prime\prime}$, i.e., the transition point between the $O_h\rightarrow D_3$ and Kitaev phases. 
The transition at $\phi_c^\prime\sim 0.11\pi$ between $O_h\rightarrow D_3$ and $O_h\rightarrow D_4$ phases is also identifiable.
Better transition values are obtained by iDMRG calculations which will be discussed in Sec. \ref{sec:Kitaev}.

\subsection{Spin expectation values}

We  further study the spin-spin correlation functions under different magnetic fields at $\phi=0.06\pi$. 

\subsubsection{Field along $(1,1,1)$-direction}

\begin{figure}
\center
\includegraphics[width=8.3cm]{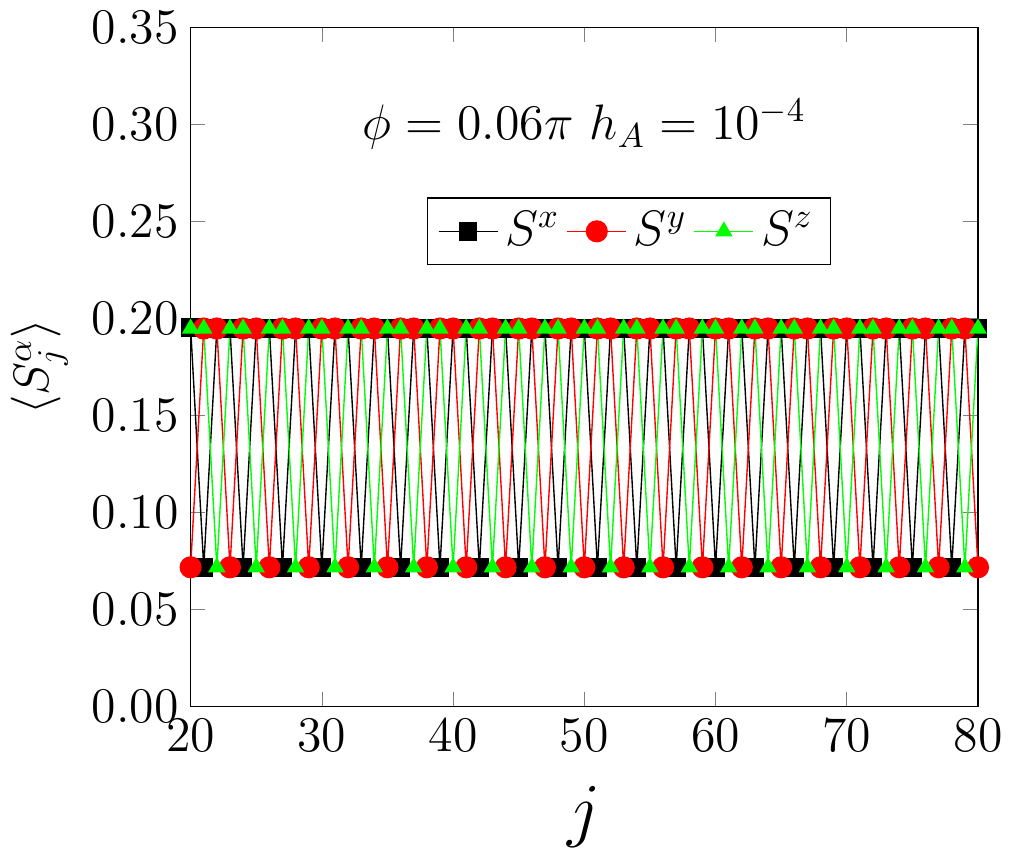}
\caption{Spin expectation values $\left<S_j^\alpha\right>$ ($j\in \mathbb{Z},\alpha=x,y,z$) with a small $10^{-4}$ field along the $(1,1,1)$-direction. 
DMRG numerics are performed on $L=96$ sites at $\phi=0.06\pi$ with periodic boundary conditions. 
} 
\label{fig:Local_Spin_111}
\end{figure}

We apply a small field $h_A=10^{-4}$ along the $(1,1,1)$-direction,
and compute the spin expectation values $\left<S_j^\alpha\right>$ ($j\in \mathbb{Z},\alpha=x,y,z$).
If the system has an $O_h\rightarrow D_3$ classical order, then
 the spin alignments should satisfy the following pattern,
\bea
\vec{S}_1=\left(\begin{array}{c}
a\\
a\\
b
\end{array}
\right),
\vec{S}_2=\left(\begin{array}{c}
a\\
b\\
a
\end{array}
\right),
\vec{S}_3=\left(\begin{array}{c}
b\\
a\\
a
\end{array}
\right).
\label{eq:spin_ori_Classical_2}
\eea
As can be seen from Fig. \ref{fig:Local_Spin_111}, this is indeed satisfied with $a\sim 0.20$, and $b\sim0.075$, which is consistent with the values determined in Eq. (\ref{eq:a2b2ab}).

\subsubsection{Field along $(0,0,1)$-direction}

\begin{figure}
\center
\includegraphics[width=8.3cm]{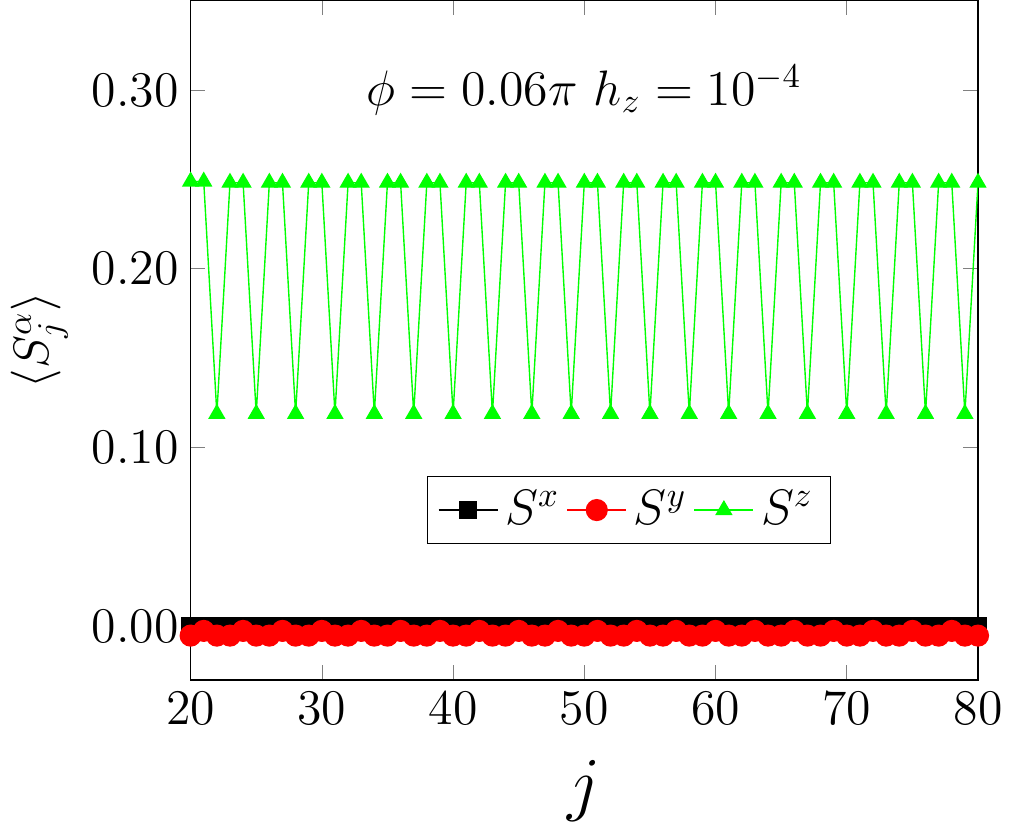}
\caption{Spin expectation values $\left<S_j^\alpha\right>$ ($j\in \mathbb{Z},\alpha=x,y,z$) with a small $10^{-4}$ field along the $(0,0,1)$-direction. 
DMRG numerics are performed on $L=96$ sites at $\phi=0.06\pi$ with periodic boundary conditions. 
} 
\label{fig:Local_Spin_001}
\end{figure}

For completeness of discussion, 
we also apply a small field $h_z=10^{-4}$ along the $(0,0,1)$-direction,
and compute the spin expectation values $\left<S_j^\alpha\right>$ ($j\in \mathbb{Z},\alpha=x,y,z$).
Obviously, if the system has an $O_h\rightarrow D_4$  order, then
 the spin alignments should satisfy the following pattern,
\bea
\vec{S}_1=\left(\begin{array}{c}
0\\
0\\
b^\prime
\end{array}
\right),
\vec{S}_2=\left(\begin{array}{c}
0\\
0\\
a^\prime
\end{array}
\right),
\vec{S}_3=\left(\begin{array}{c}
0\\
0\\
a^\prime
\end{array}
\right).
\label{eq:spin_ori_nonclassical_2}
\eea
However, less obvious is that even if the system has an $O_h\rightarrow D_3$ order, it is still possible for the spin orientations to exhibit the pattern in Eq. (\ref{eq:spin_ori_nonclassical_2}).
Consider the state
\bea
\ket{G}=\frac{1}{2}(\ket{A}+\ket{B}+\ket{C}+\ket{D}),
\eea
where $\ket{\Lambda}$ ($\Lambda=A,B,C,D$) denotes the ground state of $O_h\rightarrow D_3$ order in which the ``center of mass" directions of the spins in a unit cell points to the vertex $\Lambda$ of the cube as shown in Fig. \ref{fig:spin_orders}.
Suppose the order is $O_h\rightarrow D_3$,
then it can be checked in the presence of the $(0,0,1)$-field, 
the state $\ket{G}$ is a ground state of the system and the spin alignments are exactly given by Eq. (\ref{eq:spin_ori_nonclassical_2}).

The DMRG numerical results are displayed in Fig. \ref{fig:Local_Spin_001},
where $\phi=0.06\pi$ and $L=96$ with periodic boundary conditions.
As can be seen from Fig. \ref{fig:Local_Spin_001}, Eq. (\ref{eq:spin_ori_nonclassical_2}) is satisfied.
However, as discussed in above, this does not necessarily mean that the system has an $O_h\rightarrow D_4$ order.

\subsection{``Center of mass" spin directions}

\begin{figure}
\center
\includegraphics[width=7cm]{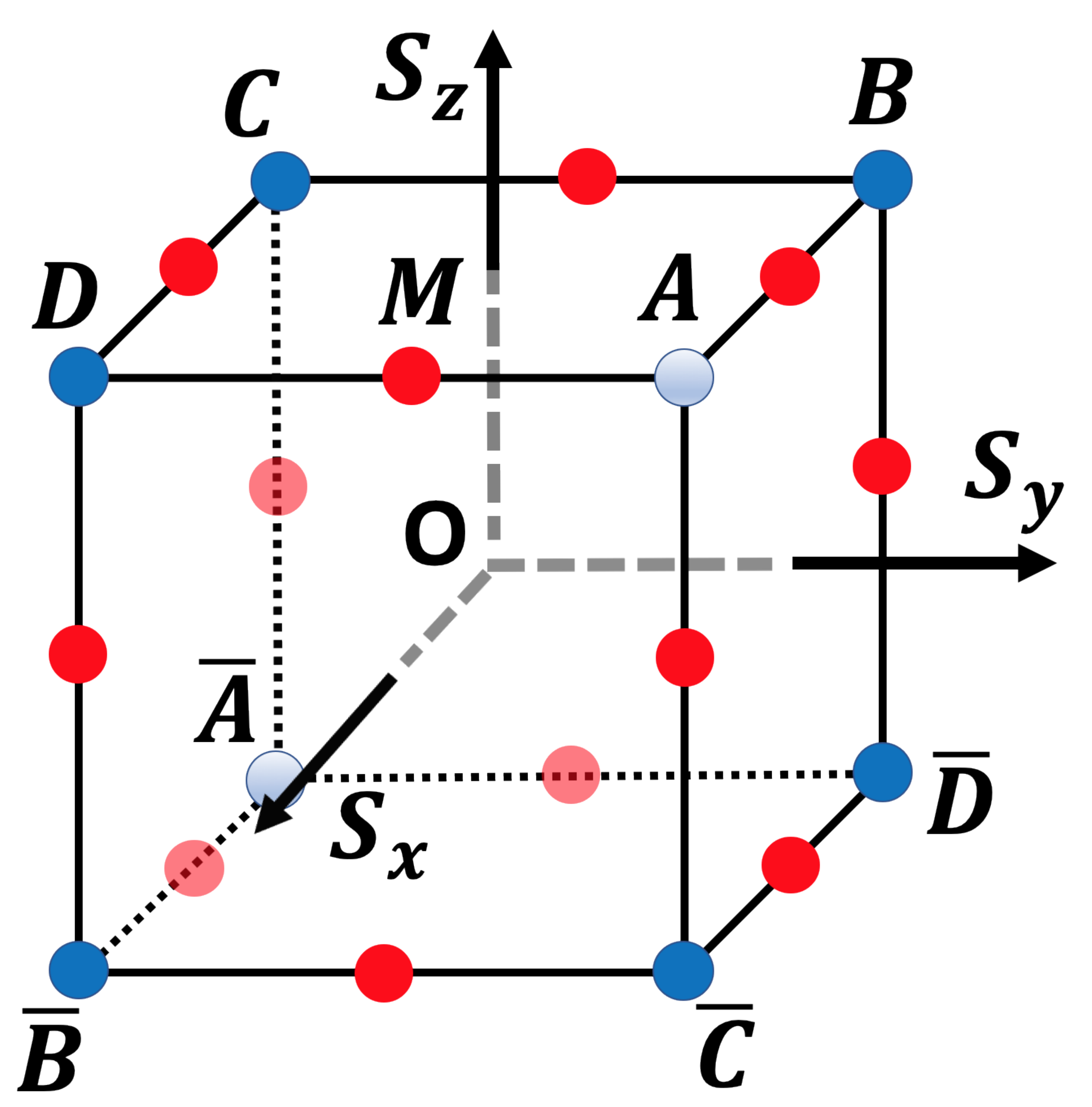}
\caption{Center of mass spin directions for the two possible rank-1 magnetic orders with 8-fold (blue circles) and 12-fold (red circles), respectively. 
} 
\label{fig:spin_orientations}
\end{figure}

Previously, we have excluded the $O_h\rightarrow D_4$ order which has six degenerate ground states.
However, there are still other possibilities. 
For example, one can imagine that the center of mass directions point to the twelve red circles in Fig. \ref{fig:spin_orientations} (referred as 12-fold magnetic order in what follows).
Therefore, we need some smoking-gun evidence for the proposed $O_h\rightarrow D_3$ order.
This is what we will do in this section.

Notice that the most prominent feature of the $O_h\rightarrow D_3$ order is that the center of mass spin directions point towards the vertices of the cube. 
Thus, a smoking-gun study is to directly check the center of mass direction.
So the strategy is to apply some small field such that the vertex $A$ in Fig. \ref{fig:spin_orientations} is selected out, and then check whether the center of mass direction is along $(1,1,1)$.

The coordinates of $A,S_z,M$ are $A=(1,1,1)$, $S_z=(0,0,1)$, and $M=(1,0,1)$, respectively. 
Let $P$ be a point lying in the triangle $AS_zM$.
Then the advantage of a $P$-field is that it is able to select out a unique state if the system has the corresponding order.
For example, if the system has $O_h\rightarrow D_3$ or $O_h\rightarrow D_4$ or 12-fold orders,
then the $P$-field selects out the states located at $A$ or $S_z$ or $M$ correspondingly.
A simple choice of $P$ would be the normalized direction of the sum of the coordinates of $A,S_z,B$, i.e.,
\bea
P=\frac{1}{\sqrt{14}}(2,1,3)^T.
\label{eq:P_field}
\eea

Define the ``center of mass" spin with a unit cell as
\bea
\vec{S}_{c}(n)=\vec{S}_{1+3n}+\vec{S}_{2+3n}+\vec{S}_{3+3n}.
\eea
Define a coordinate transformation according to 
\begin{eqnarray}
x^\prime &=& -\frac{1}{\sqrt{6}} x+\sqrt{\frac{2}{3}} y-\frac{1}{\sqrt{6}} z ,\nn\\
y^\prime &=&-\frac{1}{\sqrt{2}} x + \frac{1}{\sqrt{2}} z, \nn\\
z^\prime &=& \frac{1}{\sqrt{3}} x+\frac{1}{\sqrt{3}} y+\frac{1}{\sqrt{3}} z.
\label{eq:rotate_6rot}
\end{eqnarray}
Then it is clear that the new axis $z^\prime$ points to the $A$ vertex.
In what follows, we consider the components of $\vec{S}_c(n)$ in the new coordinate frame:
\bea
S^{x\prime}_{c}(n)&=&-\frac{1}{\sqrt{6}} S^{x}_{c}(n)+\sqrt{\frac{2}{3}} S^{y}_{c}(n)-\frac{1}{\sqrt{6}} S^{z}_{c}(n),\nn\\
S^{y\prime}_{c}(n) &=&-\frac{1}{\sqrt{2}} S^{x}_{c}(n) + \frac{1}{\sqrt{2}} S^{z}_{c}(n), \nn\\
S^{z\prime}_{c}(n) &=& \frac{1}{\sqrt{3}} S^{x}_{c}(n)+\frac{1}{\sqrt{3}} S^{y}_{c}(n)+\frac{1}{\sqrt{3}} S^{z}_{c}(n).
\label{eq:rotated_6}
\eea

\subsubsection{``Center of mass" correlation functions}

\begin{figure}
\center
\includegraphics[width=8.3cm]{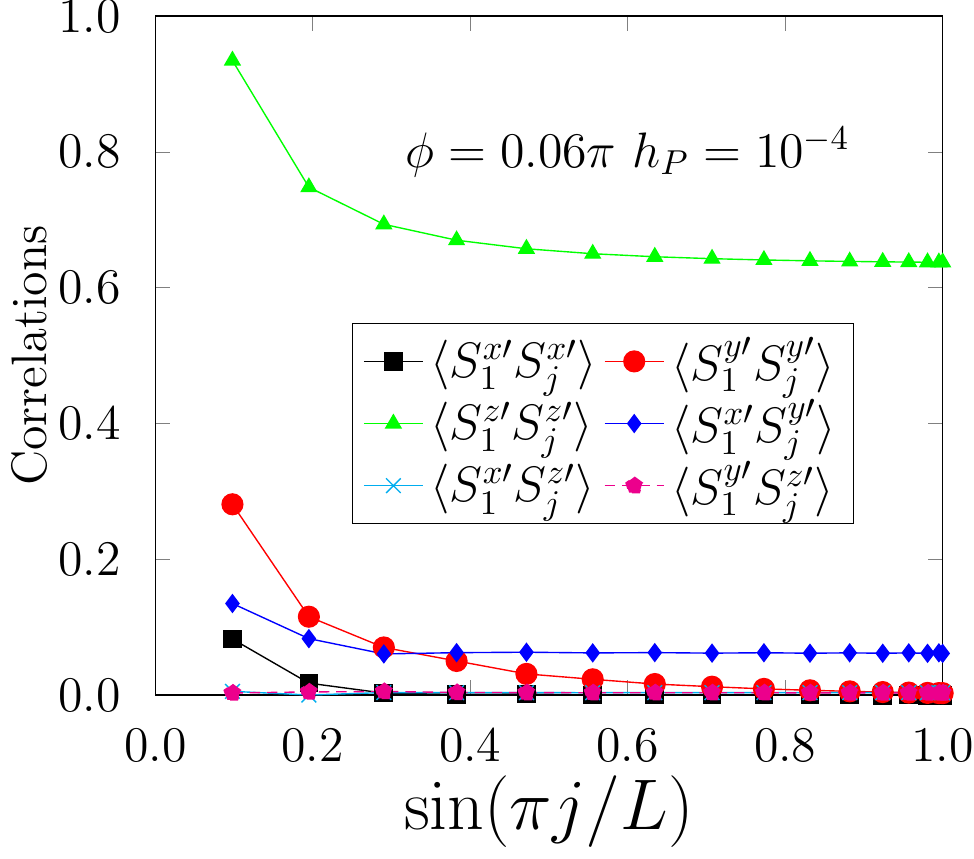}
\caption{Correlation functions of center of mass spins in the rotated frame defined by Eq. (\ref{eq:rotated_6}) under a small $(2,1,3)$-field. 
} 
\label{fig:hp_field_OhD3_CorrLinear}
\end{figure}

We apply a small $h_P$ field defined according to Eq. (\ref{eq:P_field}), i.e.,
\bea
-h_P\sum_{j} \frac{1}{\sqrt{14}} (2S_j^x+S_j^y+3S_j^z).
\label{eq:hP_field}
\eea
The field strength is taken as $h_P=10^{-4}$, and the system size is  chosen as $L=96$ in numerical calculations.
The numerical results for the correlation functions of center of mass spins are displayed in Fig. \ref{fig:hp_field_OhD3_CorrLinear}.
If the order is $O_h\rightarrow D_3$, then only $\langle  S^{z\prime}_{c}(1)S^{z\prime}_{c}(n)\rangle$ is nonzero, and all others should vanish.
Indeed, as shown by Fig. \ref{fig:hp_field_OhD3_CorrLinear}, only $\langle  S^{z\prime}_{c}(1)S^{z\prime}_{c}(n)\rangle$ is significant, and all other correlations are very small.
According to our previous analysis, this provides strong evidence for the $O_h\rightarrow D_3$ order.

\subsubsection{``Center of mass" spin expectation values}

\begin{figure}
\center
\includegraphics[width=8.3cm]{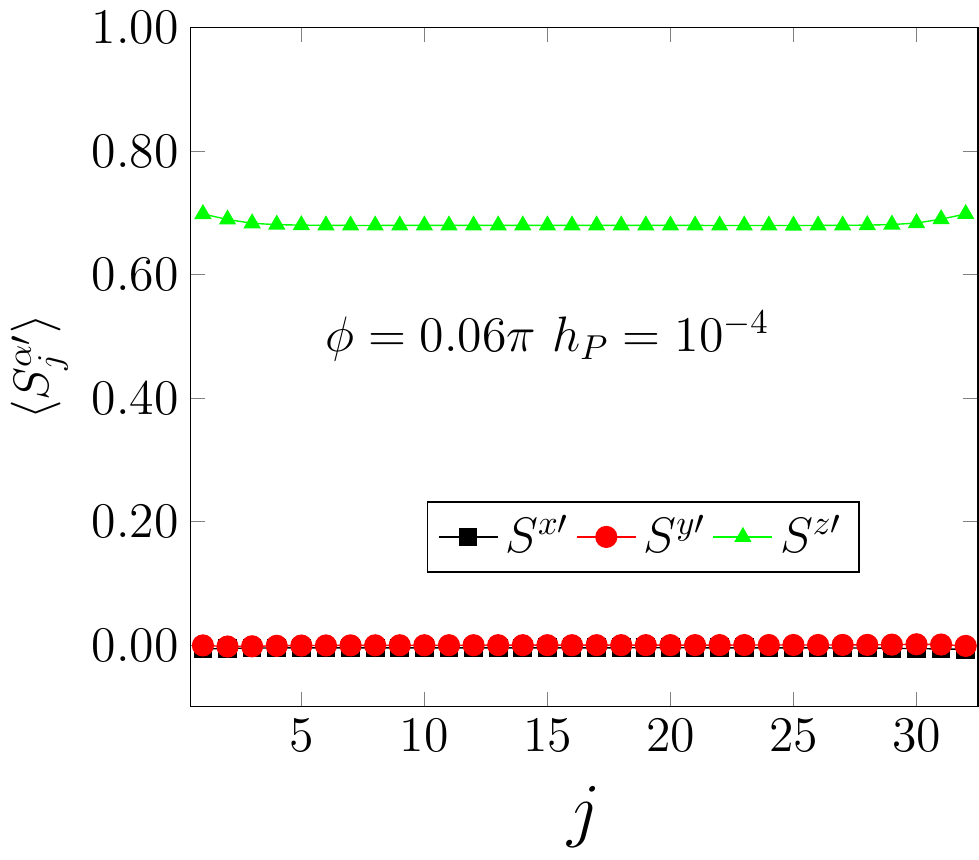}
\caption{Expectation values of center of mass spins in the rotated frame defined by Eq. (\ref{eq:rotated_6}) under a small $(2,1,3)$-field. 
} 
\label{fig:hp_field_OhD3}
\end{figure}

This time we  directly measure the spin expectation values $\langle  S^{\alpha\prime}_{c}(n)\rangle$ in the new frame.
If the order is $O_h\rightarrow D_3$, then only $\langle  S^{z\prime}_{c}(n)\rangle$ is nonzero.
As is clear from Fig. \ref{fig:hp_field_OhD3}, indeed only $\langle  S^{z\prime}_{c}(n)\rangle$ does not vanish, and both $\langle  S^{x\prime}_{c}(n)\rangle$ and $\langle  S^{y\prime}_{c}(n)\rangle$ are negligible. 

\section{Origin of classical order}
\label{sec:origin_classical}

\begin{figure}
\center
\includegraphics[width=7.5cm]{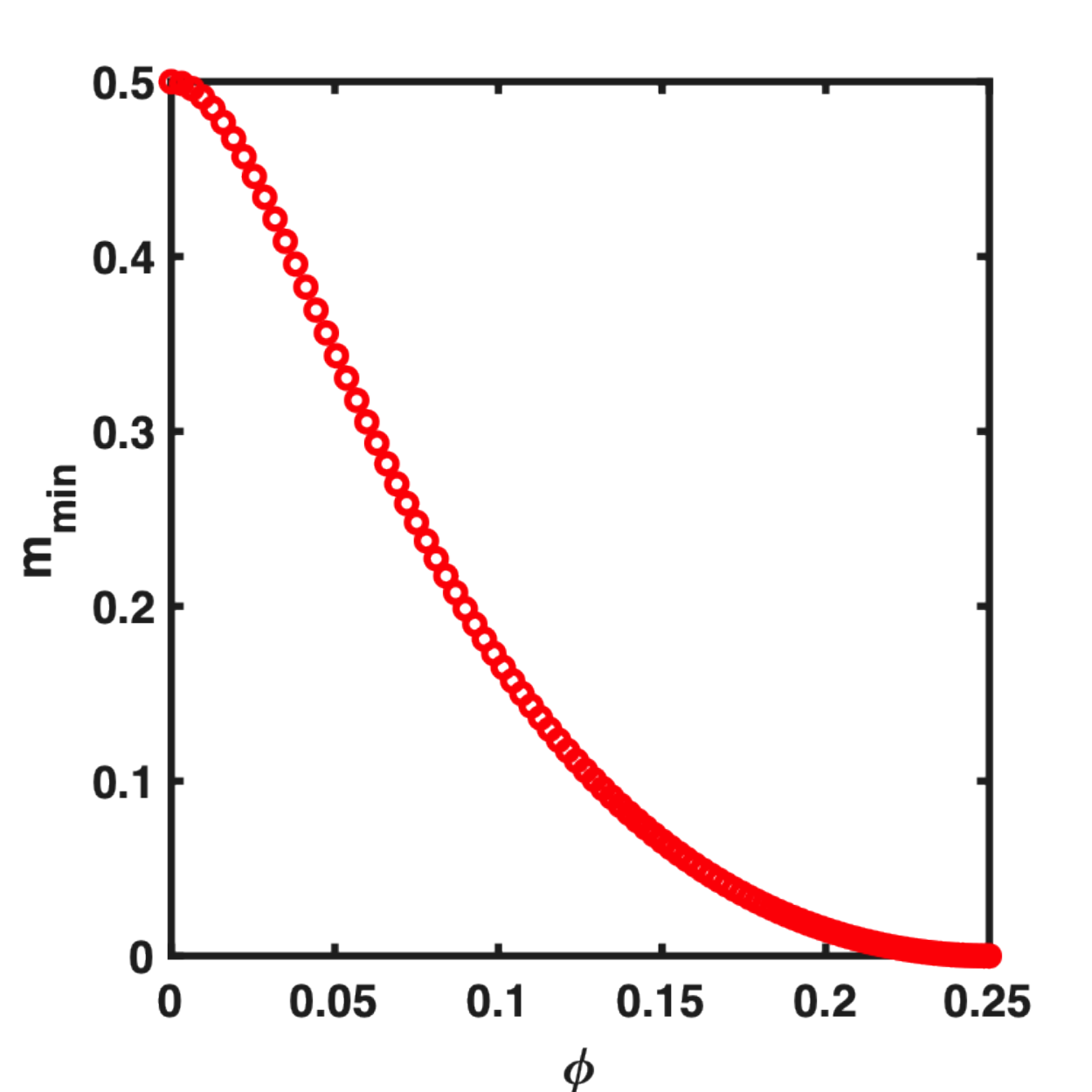}
\caption{Smallest spin wave mass as a function of $\phi$ calculated from spin wave theory in the large-$S$ limit.
} 
\label{fig:mass_vs_phi}
\end{figure}

In this section, we give a naive argument for the origin of the classical order in the region where one expects strong quantum fluctuations.  

As discussed in details in Ref. \onlinecite{Yang2020}, the system has a classical $O_h\rightarrow D_3$ order  for all spin values equal to or greater than $1$ in the vicinity of $\phi=0.25\pi$.
So spin-$1/2$ is the only exception where the order is demonstrated to be $O_h\rightarrow D_4$.
Classical analysis predicts an $O_h\rightarrow D_3$ order in the entire range $\phi\in[0,0.25\pi]$,
and the eight degenerate ground states are represented by the eight blue circles in Fig. \ref{fig:spin_orientations}.
When $S=1/2$, quantum fluctuations become strong, hence one would expect that the tunneling effects among these eight circles become important, which changes the potential minima from the eight vertices of the cube into the six coordinate directions $\pm \hat{\alpha}$ ($\alpha=x,y,z$), invalidating the classical analysis. 

Fig. \ref{fig:mass_vs_phi} displays the smallest spin wave mass $m_{\text{min}}$ calculated from the   spin wave theory within the range $[0,0.25\pi]$ using the method in Ref. \onlinecite{Yang2020}.
A feature is that $m_{\text{min}}$ increases monotonically as the angle $\phi$ decreases.
Naively, when $m_{\text{min}}$ is large, the potential minima become steeper, making the tunneling more difficult.
Thus it is possible that in the spin-$1/2$ case,  for very small $\phi$, the tunneling effects are suppressed by the larger values of the spin wave mass, and the classical order emerges. 
Of course, the $\phi$'s cannot be too small, otherwise the system goes into the other limit of $\phi$,
which is controlled by the exactly solvable  Kitaev point $\phi=0$
with an infinite ground state degeneracy \cite{Brzezicki2007}.
Such intermediate $\phi$-range turns out to be $[0.033\pi,0.1\pi]$.
Notice that the classical order arises from the competition between confinement of classical minima and quantum mechanical tunneling. 
So one would expect that the gap should be very small. 

Future analytic studies about the nonclassical origin of the classical order  are desired, including instanton calculations on the tunneling among classical configurations and higher order  $1/S$-expansion in the framework of spin wave theory.

\section{The ``Kitaev" phase}
\label{sec:Kitaev}

\begin{figure*}[htbp]
\includegraphics[width=8.5cm]{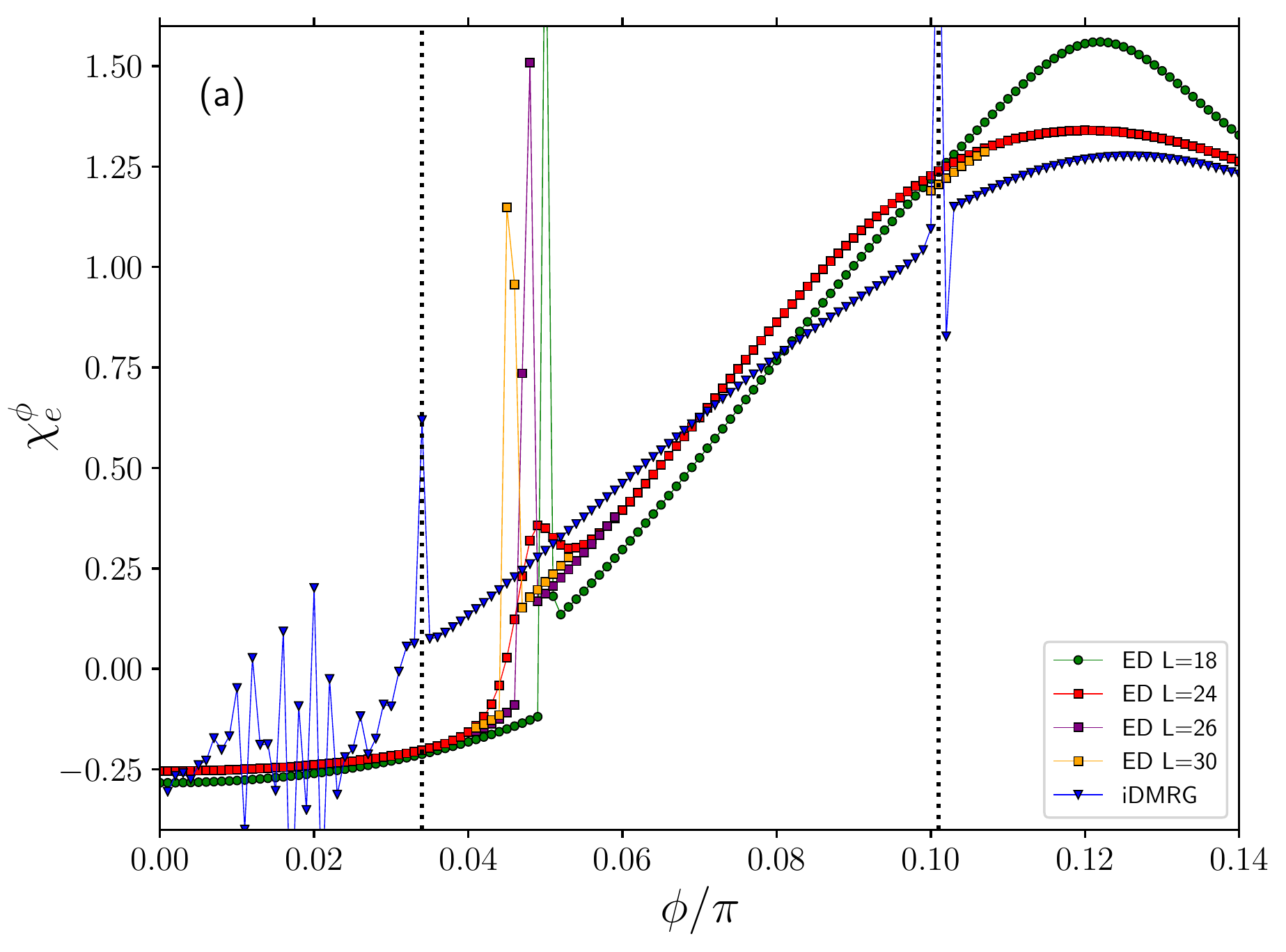}
\includegraphics[width=8.5cm]{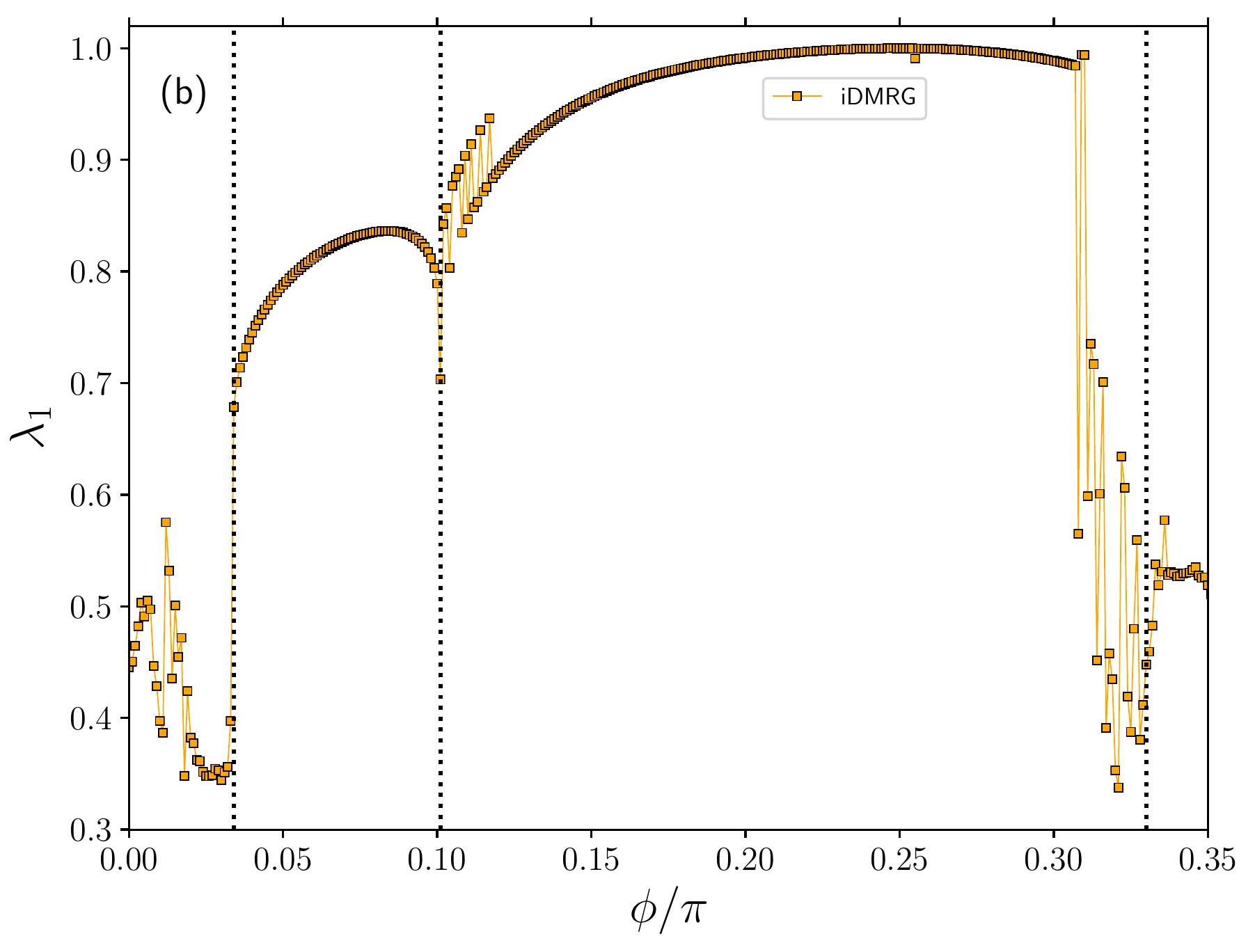}
\caption{(a) $\chi_e^\phi$ vs $\phi$ calculated from ED for $L=18,24,26,30$ sites and iDMRG,
(b) $\lambda_1$ vs $\phi$ calculated from iDMRG.
In (a,b), the calculations are performed in the original frame defined in Eq. (\ref{eq:H_unrot}).
The irregularities in the figures may come from numerical instabilities in the corresponding regions of parameters.
} \label{fig:chi}
\end{figure*}

\begin{figure*}[htbp]
\includegraphics[width=18cm]{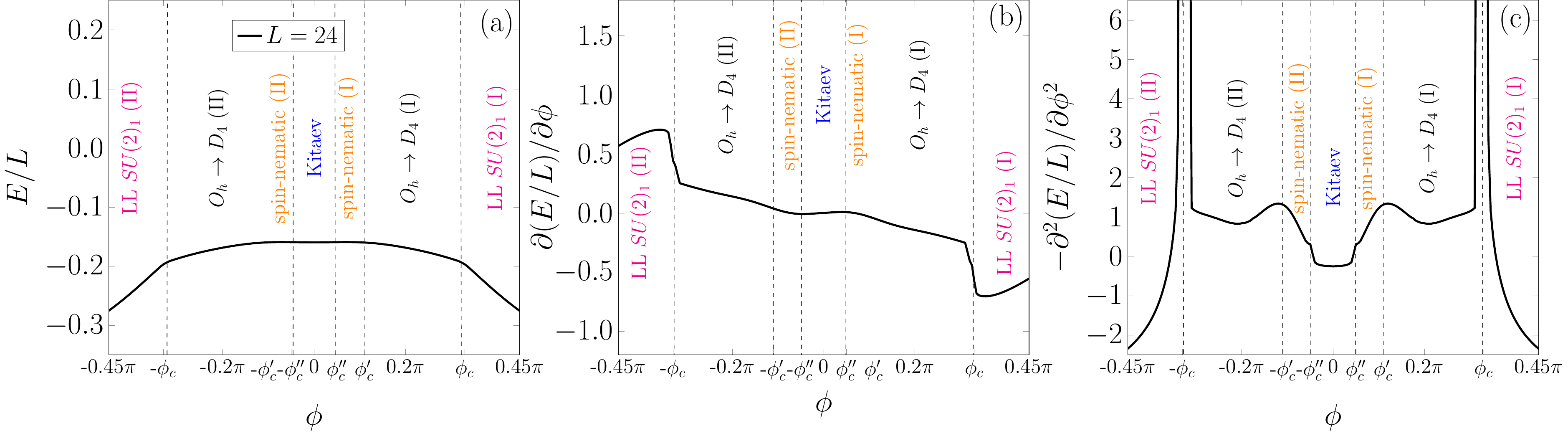}
\caption{(a) $E/L$, (b) $\partial (E/L)/\partial \phi$, and (c) $\partial^2 (E/L)/\partial \phi^2$ as functions of $\phi$.
DMRG calculations are performed on $L=24$ sites with periodic boundary conditions.
} \label{fig:ground_E}
\end{figure*}

In this section, we briefly discuss the ``Kitaev" phase in the range $\phi\in[-\phi_c^{\prime\prime},\phi_c^{\prime\prime}]$, where the physics remains unclear.
As shown in Fig. \ref{fig:ground_E}, the study of the ground state energy shows no signature of singularity at the AFM Kitaev point $\phi=0$, therefore the intervals $[-\phi_c^{\prime\prime},0]$ and $[0,\phi_c^{\prime\prime}]$ are likely in the same phase.
As can be seen from Fig. \ref{fig:chi} (a), the ED results of $\chi_e^\phi=-\partial^2 e_0/\partial \phi^2$ show big peaks  for $L=18,26,30$ sites around $\phi=0.05\pi$ (except $L=24$ where the peak is small),
where $e_0=E/L$ is the ground state energy per site.
However, it can be observed that the peak position shifts to smaller value of $\phi$ by increasing the system size.
Indeed, the iDMRG  results of $\chi_e^\phi$ in Fig. \ref{fig:chi} (a) predicts $\phi_c^{\prime\prime}$ to be $0.034\pi$, which is consistent with the sudden jump of $\lambda_1$ in Fig. \ref{fig:chi} (b) at the same value of $\phi$,
where $\lambda_1$ is the largest eigenvalue of the reduced density matrix of a subsystem of $L/2-1$ sites.
Based on this, we conjecture that there may be a strong finite size dependence of $\phi_c^{\prime\prime}$, and the thermodynamic value of $\phi_c^{\prime\prime}$ is possibly $0.034\pi$ as given by the iDMRG results in Fig. \ref{fig:chi} (a,b).

The study of the  ground state degeneracy in the Kitaev phase shows a strong finite size dependence and no reliable value can be extracted. 
In addition, we find no response to small spin-nematic nor rank-1 magnetic fields in the Kitaev phase.
The AFM Kitaev point ($K>0,\Gamma=0$) is exactly solvable through Jordan-Wigner transformation, and it is known that the ground state is $2^{L-1}$-fold degenerate for a system of finite size $L$ \cite{Brzezicki2007}. 
In the thermodynamic limit, this becomes an exponentially large infinite degeneracy.
Therefore, it is expected that there are huge quantum fluctuations in the Kitaev phase.
The irregular behaviors in the Kitaev phase in the iDMRG results in Fig. \ref{fig:chi} (a,b) possibly arise from  convergence problems due to the large number of nearly degenerate states (exactly degenerate at $\phi=0$).
Whether there exists a topological string order in the Kitaev phase remains to be explored further.

\section{Conclusions}
\label{sec:concl}

In summary, we have studied the phase diagram of the spin-1/2 Kitaev-Gamma chain with an  AFM Kitaev coupling.
In addition to the emergent SU(2)$_1$ and the $O_h\rightarrow D_4$ phases established in Ref. \onlinecite{Yang2019}, 
two new phases are identified, i.e., a phase with $O_h\rightarrow D_3$ symmetry breaking and a ``Kitaev" phase.
The $O_h\rightarrow D_3$ symmetry breaking corresponds to the classical spin order, but appears in the region very close to the AFM Kitaev point where the quantum fluctuations are presumably strong.
Furthermore, a two-step symmetry breaking $O_h\rightarrow D_{3d}\rightarrow D_3$ is observed as the length scale is increased.
For the ``Kitaev" phase, no evidence of any spin ordering nor Luttinger liquid behavior is found,
and its nature remains unclear. 
Whether there exists any topological string order \cite{Catuneanu2019} in the  ``Kitaev" phase is worth further studies.

{\it Acknowledgments}
WY and IA acknowledge support from NSERC Discovery Grant 04033-2016.
HYK acknowledges support from NSERC Discovery Grant 06089-2016, the Centre for Quantum
Materials at the University of Toronto and the Canadian Institute for Advanced Research.
AN acknowledges computational resources and services provided by Compute Canada and
Advanced Research Computing at the University of British Columbia.
AN is supported by the Canada First Research Excellence Fund.
ESS acknowledges support from NSERC Discovery Grant,  SHARCNET (www.sharcnet.ca), and Compute/Calcul Canada (www.computecanada.ca).

\appendix 

\begin{widetext}

\section{Explicit expressions of the Hamiltonian}
\label{sec:Ham}

In this appendix, we spell out the terms in the Hamiltonians in different frames. 
In general, we write the Hamiltonian $H$ as $H=\sum_{j=1}^L H_{j,j+1}$ where $H_{j,j+1}$ is the term on the bond between the sites $j$ and $j+1$.   
The forms of $H_{j,j+1}$ will be written explicitly. 

In the unrotated frame, the form of $H_{j,j+1}$ has a two-site periodicity. 
We have
\bea
H_{2n+1,2n+2}&=&K S_{2n+1}^x S_{2n+2}^x +\Gamma (S_{2n+1}^y S_{2n+2}^z+S_{2n+1}^z S_{2n+2}^y), \nn\\
H_{2n+2,2n+3}&=&K S_{2n+2}^y S_{2n+3}^y +\Gamma (S_{2n+2}^z S_{2n+3}^x+S_{2n+2}^x S_{2n+3}^z).
\eea

In the six-sublattice rotated frame, the form of $H_{j,j+1}$ has a three-site periodicity. 
We have
\bea
H^\prime_{3n+1,3n+2}&=&-KS_{3n+1}^xS_{3n+2}^x-\Gamma (S_{3n+1}^yS_{3n+2}^y+S_{3n+1}^zS_{3n+2}^z),\nn\\
H^\prime_{3n+2,3n+3}&=&-KS_{3n+2}^zS_{3n+3}^z-\Gamma (S_{3n+2}^xS_{3n+3}^x+S_{3n+2}^yS_{3n+3}^y),\nn\\
H^\prime_{3n+3,3n+4}&=&-KS_{3n+3}^yS_{3n+4}^y-\Gamma (S_{3n+3}^zS_{3n+4}^z+S_{3n+3}^xS_{3n+4}^x).
\eea

\section{Spin-nematic order parameters}
\label{app:spin_nematic}

In this appendix, we derive the spin-nematic order parameters by requiring the $O_h\rightarrow D_{3d}$ symmetry breaking.

Define a $9\times 9$ matrix $M$ as
\bea
M=\bra{\Omega_e} \left(\begin{array}{c}
\vec{S}_1\\
\vec{S}_2\\
\vec{S}_3
\end{array}
\right)
(\vec{S}_1^T~\vec{S}_2^T~\vec{S}_3^T)
\ket{\Omega_e},
\label{eq:M}
\eea
in which $\ket{\Omega_e}$ is one of the four symmetry breaking ground states,
and $\vec{S}_i=(S^x_i,S^y_i,S^z_i)^T$ is viewed as a three-component column vector.
In Eq. (\ref{eq:M}), the site indices should be understood as modulo $3$ such that the expectation values are taken for adjacent sites.
For example, $S_3^\alpha S_1^\beta$ means $S_{3+3n}^\alpha S_{4+3n}^\beta$
where $n\in\mathbb{Z}$.
We note that the value of $n$ is not essential in Eq. (\ref{eq:M}) since $T_{3a}$ is assumed to be unbroken.
Also notice that $M$ includes all possible expectation values of adjacent-site  spin-nematic order parameters.

Before proceeding on, we give the explicit expression of $D_{3d}$.
Assuming the unbroken symmetry group to be 
\bea
G_1=\mathopen{<}\mathcal{T},R_aT_a,R_II,T_{3a}\mathclose{>},
\eea
we will show that $G_1/\mathopen{<}\mathcal{T},T_{3a}\mathclose{>}$ is isomorphic to $D_3$.
Hence, $D_{3d}$ is given by
\bea
D_{3d}=\mathopen{<}\mathcal{T},R_aT_a,R_II\mathclose{>}/\mathopen{<}T_{3a}\mathclose{>}.
\eea

The isomorphism $G_1/\mathopen{<}\mathcal{T},T_{3a}\mathclose{>}\simeq  D_3$ can be proved
using the following generator-relation representation of $D_{n}$:
\bea
D_n=\mathopen{<} a,b| a^n=b^2=(ab)^2=e \mathclose{>},
\label{eq:generator_Dn}
\eea
in which $e$ is the identity element. 
Define $a=R_aT_a$, $b=R_I I$.
Then $a^3=T_a^3$, and $b^2=e$, which are both equal to the identity element modulo $T_{3a}$. 
Hence $G_1/\mathopen{<}T,T_a \mathclose{>}$ is a subgroup of $D_3$ since it is generated by $\{R_aT_a,R_I I\}$.
On the other hand, $\{e,R_a,R_a^{-1},R_I, R_I^{-1},R_aR_I\}$ are all distinct operations,
and are the actions of the elements of $G_1/\mathopen{<}\mathcal{T},T_a \mathclose{>}$ restricted within the spin space.
This shows that there are at least six elements in $G_1/\mathopen{<}\mathcal{T},T_a \mathclose{>}$.
But the order of $D_3$ is six, thus $G_1/\mathopen{<}\mathcal{T},T_a \mathclose{>}$ is isomorphic to $D_3$.

Having $D_{3d} $ at hand, 
the next step is to solve the most general form of $M$ in Eq. (\ref{eq:M})
by assuming the $D_{3d}$ invariance of $\ket{\Omega_e}$.
Since the spin-nematic order parameters automatically maintain the time reversal symmetry,
 $T$ has no restriction on the form of $M$
and it is enough to consider $R_aT_a$ and $R_I I$.
Using
\bea
R_aT_a (\vec{S}_1^T~\vec{S}_2^T~\vec{S}_3^T) (R_aT_a)^{-1}
&=&(\vec{S}_1^T~\vec{S}_2^T~\vec{S}_3^T) U_a,\nn\\
R_I I (\vec{S}_1^T~\vec{S}_2^T~\vec{S}_3^T) (R_II)^{-1}
&=&(\vec{S}_1^T~\vec{S}_2^T~\vec{S}_3^T) U_I,
\eea
we obtain the constraints on $M$ as 
\bea
U_a M  U_a^{-1}=M,
\label{eq:Eq_M_a}
\eea
and
\bea
U_I M  U_I^{-1}=M,
\label{eq:Eq_M_I}
\eea
in which 
\bea
U_a=\left(\begin{array}{ccc}
0&0&R_a\\
R_a&0&0\\
0&R_a&0
\end{array}\right),~~
U_I=\left(\begin{array}{ccc}
0&0&R_I\\
0&R_I&0\\
R_I&0&0
\end{array}\right),
\label{eq:UaUI}
\eea
where 
\bea
R_a=\left(\begin{array}{ccc}
0&1&0\\
0&0&1\\
1&0&0
\end{array}\right),~~
R_I=\left(\begin{array}{ccc}
0&0&-1\\
0&-1&0\\
-1&0&0
\end{array}\right).
\label{eq:RaRI}
\eea

The requirement Eq. (\ref{eq:Eq_M_a}) leads to 
\bea
M=\left(\begin{array}{ccc}
A & C^T & R_a^{-1}CR_a\\
C & R_aAR_a^{-1}& R_aC^TR_a^{-1}\\
R_a^{-1} C^TR_a & R_aCR_a^{-1}& R_a^{-1}AR_a
\end{array}
\right),
\eea
in which $A$ is a symmetric matrix.
Eq. (\ref{eq:Eq_M_a}) put further constraints on $A$ and $C$ as
\bea
(R_aR_I) A (R_aR_I)^{-1} &=& A\nn\\
(R_aR_I) C (R_aR_I)^{-1} &=& C^T.
\label{eq:Relations_A_C}
\eea

Next we solve all possible forms of $A$ and $C$ satisfying Eq. (\ref{eq:Relations_A_C}).
Using $R_a$ and $R_I$ given in Eq. (\ref{eq:RaRI}), we are able to obtain
\bea
A=\left(\begin{array}{ccc}
\lambda&\nu&\sigma\\
\nu&\lambda&\sigma\\
\sigma&\sigma&\mu
\end{array}
\right),~
C=\left(\begin{array}{ccc}
a&c&d\\
d&b&f\\
c&e&b
\end{array}
\right).
\eea

As a result, there are ten linear independent solutions of $M$, summarized as follows,
\bea
\lambda&=&\left<S_1^xS_1^x\right>=\left<S_1^yS_1^y\right>=\left<S_2^xS_2^x\right>
=\left<S_2^zS_2^z\right>=\left<S_3^yS_3^y\right>=\left<S_3^zS_3^z\right>\nn\\
\mu&=&\left<S_1^zS_1^z\right>=\left<S_2^yS_2^y\right>=\left<S_3^xS_3^x\right>\nn\\
\nu&=&\left<S_1^xS_1^y\right>=\left<S_1^yS_1^x\right>=\left<S_2^xS_2^z\right>
=\left<S_2^zS_2^x\right>=\left<S_3^yS_3^z\right>=\left<S_3^zS_3^y\right>\nn\\
\sigma&=&\left<S_1^xS_1^z\right>=\left<S_1^zS_1^x\right>=\left<S_1^yS_1^z\right>
=\left<S_1^zS_1^y\right>\nn\\
&=&\left<S_2^xS_2^y\right>=\left<S_2^yS_2^x\right>=\left<S_2^yS_2^z\right>
=\left<S_2^zS_2^y\right>\nn\\
&=&\left<S_3^xS_3^y\right>=\left<S_3^yS_3^x\right>=\left<S_3^xS_3^z\right>
=\left<S_3^zS_3^x\right>,
\eea
and 
\bea
a&=&\left<S_1^xS_2^x\right>=\left<S_2^zS_3^z\right>=\left<S_3^yS_4^y\right>\nn\\
b&=&\left<S_1^yS_2^y\right>=\left<S_1^zS_2^z\right>=\left<S_2^xS_3^x\right>=\left<S_2^yS_3^y\right>=\left<S_3^xS_4^x\right>
=\left<S_3^zS_4^z\right>\nn\\
c&=&\left< S_1^x S_2^z \right>= \left< S_1^y S_2^x \right> =\left< S_2^x S_3^z \right>=\left< S_2^z S_3^y \right>=\left< S_3^yS_4^x \right> = \left< S_3^z S_4^y \right>\nn\\
d&=& \left< S_1^x S_2^y \right> =\left<S_1^z S_2^x \right>=\left< S_2^y S_3^z \right>=\left< S_2^z S_3^x \right>=\left<S_3^x S_4^y \right>=\left< S_3^y S_4^z\right>\nn\\
e&=& \left< S_1^y S_2^z \right> =\left< S_2^x S_3^y \right>= \left< S_3^z S_4^x \right> \nn\\
f&=& \left< S_1^z S_2^y \right> =\left< S_2^y S_3^x \right>= \left< S_3^x S_4^z \right>.
\label{eq:order_e}
\eea
Among these ten solutions, the first four $\lambda,\mu,\nu,\sigma$ are on-site, which we'll ignore.
In the remaining six solutions, $a$ and $b$ are just the Kitaev and Gamma couplings which are invariant under $O_h$, not just $D_{3d}$, which we also ignore.
Hence, the only relevant spin-nematic orders 
are given by $c,d,e,f$.

The spin-nematic orders $\hat{Q}_\lambda$ ($\lambda=c,d,e,f$) can be constructed by summing up the corresponding operators in Eq. (\ref{eq:quadrupole_order_c},\ref{eq:quadrupole_order_d},\ref{eq:quadrupole_order_e},\ref{eq:quadrupole_order_f}).
For example, $\hat{Q}_e$ is given by
\begin{flalign}
\hat{Q}_e=\frac{1}{L}\sum_{n}( S_{1+3n}^y S_{2+3n}^z+S_{2+3n}^x S_{3+3n}^y+S_{3+3n}^z S_{4+3n}^x),
\label{eq:Qe}
\end{flalign}
and the other three spin-nematic orders can be obtained similarly.
There are three other degenerate ground states $\ket{\Omega_{\alpha}}$ ($\alpha=x,y,z$) which can be obtained from $\ket{\Omega_e}$ by
\bea
\ket{\Omega_x}&=&R(\hat{x},\pi)\ket{\Omega_e}, \nn\\ 
\ket{\Omega_y}&=&R(\hat{y},\pi)\ket{\Omega_e},\nn\\
\ket{\Omega_z}&=&R(\hat{z},\pi)\ket{\Omega_e},
\eea
where $R(\hat{\alpha},\pi)$ ($\alpha=x,y,z$) are the representative operations in the three out of four equivalent classes in $O_h/D_{3d}$ excluding the equivalent class containing the identity element.

It is also interesting to work out the explicit forms of the spin-nematic orders within the original frame. 
The spin-nematic orders in the original frame can be obtained straightforwardly by applying the inverse of the six-sublattice rotation to the expressions in Eq. (\ref{eq:quadrupole_order_c},\ref{eq:quadrupole_order_d},\ref{eq:quadrupole_order_e},\ref{eq:quadrupole_order_f}).
The spin-nematic orders thus obtained are summarized as follows,
\bea
\hat{Q}_c^{(0)}&=&\frac{1}{L}\sum_j ( S_j^xS_{j+1}^y+S_j^yS_{j+1}^x ),\nn\\
\hat{Q}_d^{(0)}&=&\frac{1}{L}\sum_n ( S_{1+2n}^xS_{2+2n}^z+S_{1+2n}^zS_{2+2n}^x+ S_{2+2n}^yS_{3+2n}^z+S_{2+2n}^zS_{3+2n}^y ),\nn\\
\hat{Q}_e^{(0)}&=&\frac{1}{L}\sum_n ( S_{1+2n}^yS_{2+2n}^y+S_{2+2n}^xS_{3+2n}^x ),\nn\\
\hat{Q}_f^{(0)}&=&\frac{1}{L}\sum_j  S_j^zS_{j+1}^z,
\eea
in which all the spin operators refer to the original frame.

\section{Invariant correlation functions in the $O_h\rightarrow D_4$ phase}
\label{sec:inv_corr}

In this appendix, we construct the invariant correlation functions in the $O_h\rightarrow D_4$ phase.
Before proceeding to the constructions of the invariant correlation functions,
we first make some comments on the symmetry breaking pattern.
There are six equivalent classes in the quotient $O_h/D_4$, which is not a group since $D_4$ is not a normal group of $O_h$.
The six representative elements in the equivalent classes can be chosen as the group elements in $D_3=\mathopen{<}R_aT_a, R_I I\mathclose{>},\mod T_{3a}$.
Notice that this is intuitively correct since $R_a$ and $R_I$ is able to rotate the $+\hat{z}$-direction to the other five directions within $\pm \hat{\alpha}$ ($\alpha=x,y,z$).

Next consider  the correlation function $\langle S_i^\alpha S_j^\beta\rangle$.
In what follows, we will write $i,j$ to be modulo $3$, but always bear in mind that $|i-j|\rightarrow \infty$.
All the two point correlation functions are encoded in the following operators $\hat{\Phi}$,
\bea
\hat{\Phi}=\hat{S}^T \Phi \hat{S},
\label{eq:Phi_op}
\eea
in which 
\bea
\hat{S}=(S_1^x~S_1^y~S_1^z~S_2^x~S_2^y~S_2^z~S_3^x~S_3^y~S_3^z)^T,
\eea
and $\Phi$ is a $9\times 9$ numerical matrix. 
The $81$ independent correlation functions correspond to the $81$ choices of the matrix $\Phi$.

Let $\mathcal{U}\in D_3$, and $\hat{\mathcal{U}}$ be the corresponding operator in the Hilbert space.
Let $U$ be a $9\times 9$ orthogonal matrix defined as
\bea
\hat{\mathcal{U}}^{-1} \hat{S}\hat{\mathcal{U}}=U\hat{S}.
\label{eq:Transform_S}
\eea
Let $\ket{\Omega_z}$ be the ground state with all spins pointing to the $+\hat{z}$-direction.
Then the other degenerate ground states can be obtained from $\hat{\mathcal{U}}\ket{\Omega_z}$.
The invariance of the correlation function requires
\bea
\bra{\Omega_z}\hat{\mathcal{U}}^\dagger \hat{\Phi} \hat{\mathcal{U}}\ket{\Omega_z}=\bra{\Omega_z} \hat{\Phi} \ket{\Omega_z}.
\eea
Using Eq. (\ref{eq:Phi_op}) and Eq. (\ref{eq:Transform_S}), we obtain
\bea
\bra{\Omega_z} \hat{S}^T U^T\Phi U \hat{S}\ket{\Omega_z}=\bra{\Omega_z}\hat{S}^T \Phi \hat{S}\ket{\Omega_z},
\eea
which is satisfied if 
\bea
U^T\Phi U=\Phi.
\label{eq:Eq4Phi}
\eea

Since $\mathcal{U}\in D_3$, it is enough to choose the two generators $R_aT_a$ and $R_I I$ of $D_3$.
The corresponding matrix $U_a,U_I$ of these two generators have already been given in Eq. (\ref{eq:UaUI}).
Therefore, we see that Eq. (\ref{eq:Eq4Phi}) is exactly the same as Eq. (\ref{eq:Eq_M_a}) for $R_aT_a$ and as Eq. (\ref{eq:Eq_M_I}) for $R_II$.
Thus the solutions of $\Phi$ are just the same as those of $M$ in Appendix \ref{app:spin_nematic}.

In summary, the ten invariant correlation functions are
\bea
a^2+2b^2&=&\vec{S}_1\cdot \vec{S}_1+\vec{S}_2\cdot \vec{S}_2+\vec{S}_3\cdot \vec{S}_3,\nn\\
a^2&=&S_1^zS_1^z+S_2^yS_2^y+S_3^xS_3^x,\nn\\
0&=&(S_1^xS_1^y+S_1^yS_1^x)+(S_2^xS_2^z+S_2^zS_2^x)+(S_3^yS_3^z+S_3^zS_3^y),\nn\\
0&=&(S_1^xS_1^z+S_1^zS_1^x+S_1^yS_1^z+S_1^zS_1^y)+(S_2^yS_2^z+S_2^zS_2^y+S_2^xS_2^y+S_2^yS_2^x)+(S_3^xS_3^z+S_3^zS_3^x+S_3^xS_3^y+S_3^yS_3^x),\nn\\
b^2&=&S^x_1S^x_2+S^y_1S^y_3+S^z_2S^z_3,\nn\\
2ab&=&(S^y_1S^y_2+S^z_1S^z_2)+(S^x_1S^x_3+S^z_1S^z_3)+(S^x_2S^x_3+S^y_2S^y_3),\nn\\
0&=&(S_1^x S_2^z+S_1^yS_2^x)+(S_1^xS_3^y+S_1^yS_3^z)+(S_2^xS_3^z+S_2^zS_3^y),\nn\\
0&=&(S_1^xS_2^y+S_1^zS_2^x)+(S_1^yS_3^x+S_1^zS_3^y)+(S_2^yS_3^z+S_2^zS_3^x),\nn\\
0&=&S_1^yS_2^z+S_1^xS_3^z+S_2^xS_3^y,\nn\\
0&=&S_1^zS_2^y+S_1^zS_3^x+S_2^yS_3^x,
\eea
in which the symbols $\langle\cdot\cdot\cdot\rangle$ are omitted on the right hand sides of the equations.

\section{Inconsistency of four-fold  ground state degeneracy with rank-1 orders}
\label{app:inconsistency}

In this appendix, we prove that the four-fold ground state degeneracy is not consistent with any rank-1 magnetic order.
The translation symmetry by three sites is assumed to be not broken, and all symmetry operations will be considered modulo $T_{3a}$.
Therefore the full symmetry group will be referred as $O_h$ rather than $O_h\ltimes 3\mathbb{Z}$.

Suppose the unbroken symmetry group to be $D$, so that the symmetry breaking pattern is $O_h\rightarrow D$.
To obtain a four-fold degeneracy, the order of $D$ must be 12.
On the other hand, the only two subgroups of $O_h$ that have 12 elements are
the tetrahedral cubic point group $T$ and the tetragonal point group $D_{3d}$.

We show that $S_i^\alpha$ cannot be an order parameter that has either $T$ or $D_{3d}$ to be the unbroken symmetry group.
Otherwise, suppose $S_i^\alpha$ acquires a nonzero expectation value on
one of the four degenerate ground states $\ket{\Omega}$ which is assumed to be invariant under the cubic point group $T$.
Since $R(\hat{\alpha},\pi)$ ($\alpha=x,y,z$) belongs to $T$,
 the sign of $S_i^\alpha$ can be changed using $R(\hat{\beta},\pi)$ where $\beta\neq \alpha$.
As a result,
\bea
\bra{\Omega} S_i^\alpha \ket{\Omega}=-\bra{\Omega}R^{-1}(\beta,\pi) S_i^\alpha R(\hat{\beta},\pi)\ket{\Omega}.
\eea
However, since $\ket{\Omega}$ is invariant under $R(\hat{\beta},\pi)$ by assumption,
we conclude that $\bra{\Omega} S_i^\alpha \ket{\Omega}=-\bra{\Omega} S_i^\alpha \ket{\Omega}$,
which contradicts with $\bra{\Omega} S_i^\alpha \ket{\Omega}\neq 0$.
Thus, $T$ cannot be the unbroken symmetry group.
Next we consider the possibility of  $O_h\rightarrow D_{3d}$.
In the cubic group language, $D_{3d}$ contains the inversion operation. 
In our case, the time reversal operation $\mathcal{T}$ plays the role of ``inversion" when acting on spin operators since $T$ changes the sign of $\vec{S}_i$.
Since the time reversal operation belongs to $D_{3d}$,
it is clear that $D_{3d}$ again cannot be the unbroken symmetry group
because  $\vec{S}_i$ is odd under time reversal.

\section{Strength of spin-nematic orders}
\label{app:strength_sn}

\begin{figure}[h]
\includegraphics[width=7.5cm]{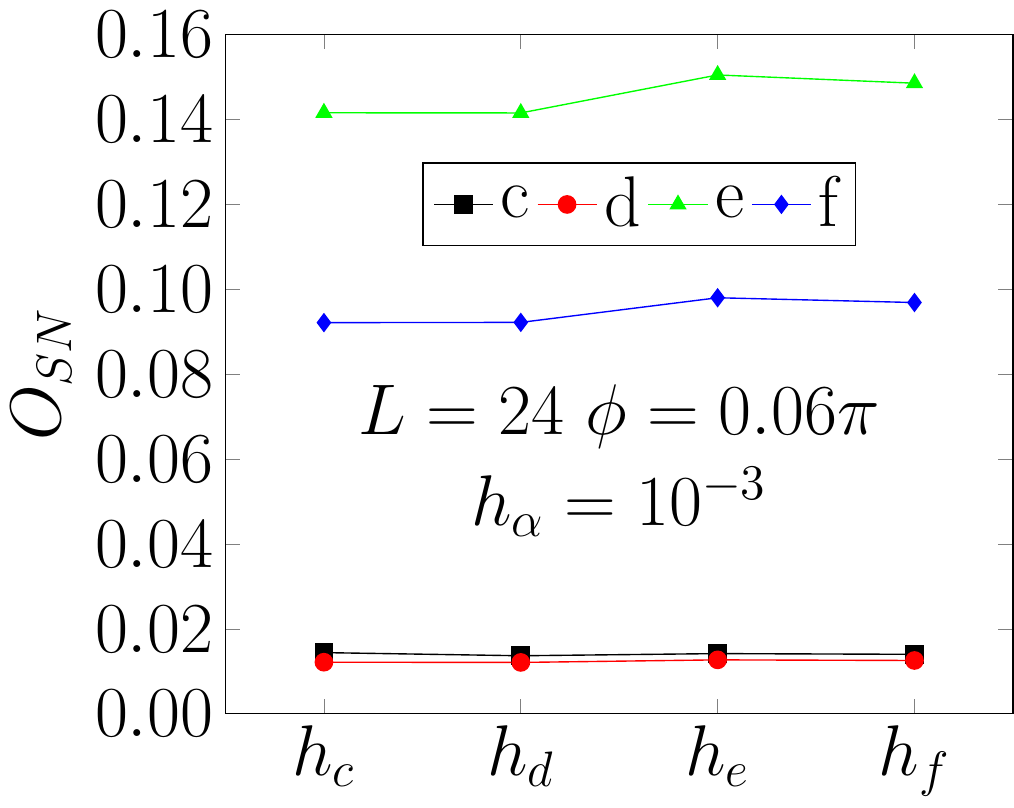}
\caption{Measured spin-nematic order parameters $c$ (black) $d$ (red), $e$ (green), $f$ (blue) under small spin-nematic fields.
The vertical axis is the measured value, and the results of different fields are displayed at different horizontal coordinates.
All the spin-nematic fields $h_c,h_d,h_e,h_f$ are taken positive with a magnitude equal to $10^{-3}$.
In all cases, ED calculations are performed on a periodic system of $L=24$ sites at $\phi=0.06\pi$.
} \label{fig:quadru_w_fields}
\end{figure}

In this appendix, we determine the values of the spin-nematic orders $c,d,e,f$ on a system of $L=24$ sites.

We note that with a field $h_e$ satisfying $\text{sgn}(h_e)=\text{sgn}(e)$,
the state $\ket{\Omega_e}$ is selected as the ground state out of the initially four nearly degenerate ground states.
Hence, we can directly compute the expectation value of the spin-nematic order parameters 
$S_{1+3n}^y S_{2+3n}^z,S_{2+3n}^x S_{3+3n}^y,S_{3+3n}^z S_{4+3n}^x$ (as given by Eq. (\ref{eq:quadrupole_order_e})) in numerics.
Notice that this cannot be done at zero spin-nematic fields since the true ground state in a finite size system may be an arbitrary linear combination of the four states $\ket{\Omega_a}$ ($a=e,x,y,z$),
which leads to a random cancellation of the expectation value due to the sign differences in Eq. (\ref{eq:quadrupole_order_e}) and Eq. (\ref{eq:e_xyz}).

We have  measured the expectation values of the spin-nematic orders under positive spin-nematic fields, and the results are shown in Fig. \ref{fig:quadru_w_fields}.
It can be read from Fig. \ref{fig:quadru_w_fields} that the expectation values are 
\bea
c\simeq 0.014, ~d\simeq 0.012,~ e\simeq 0.145,~ f\simeq 0.094, 
\label{eq:cdef1}
\eea
regardless of which field $h_\alpha$ ($\alpha=c,d,e,f$) is applied.
Here we note that as can be seen from Fig. \ref{fig:quadru_w_fields},
while the values of $c$ and $d$ are independent of $h_\lambda$ ($\lambda=c,d,e,f$),
there are small variations of $e$ and $f$ under different types of spin-nematic fields.
The reason is the same as before.
In fact, a field of $10^{-3}$ is too large for $h_e$ and $h_f$,
which mixes the ground state subspace with the excited states.
As a result, in addition to the ordering in the ground state, the order paramters also acquires contributions from excited states due to a nonzero spin-nematic susceptibility.
This explains why the measured values of $e$ and $f$ under $h_e,h_f$ are larger than those under $h_c,h_d$.

\end{widetext}


\end{document}